\documentclass[a4paper,fleqn,usenatbib]{mnras}

\usepackage{newtxtext,newtxmath}
\usepackage[T1]{fontenc}
\usepackage{ae,aecompl}
\usepackage{graphicx}
\usepackage{amsmath}
\usepackage{amssymb}
\usepackage{adjustbox}

\newcommand{\aref}[1]{\hyperref[#1]{Appendix~\ref{#1}}}

\newcommand*\samethanks[1][\value{footnote}]{\footnotemark[#1]}

\newcommand{\Msun}{\mbox{M$_{\odot}$}}

\newcommand{\HST}{\mbox{\textit{HST}}}

\newcommand{\EBVs}{\mbox{$E(B-V)_{\text{stars}}$}} 
\newcommand{\HA}{\mbox{H$\alpha$}}
\newcommand{\HB}{\mbox{H$\beta$}}
\newcommand{\HIRAC}{\mbox{$H_{160}$--IRAC}}

\newcommand{\nsamp}{350}
\newcommand{\insamp}{830}
\newcommand{\zmin}{1.25}
\newcommand{\zmax}{2.66}
\newcommand{\izmin}{1.24}

\title[Voronoi Binning Method]{The MOSDEF Survey: An Improved Voronoi Binning Technique on Spatially Resolved Stellar Populations at $z\sim2$\thanks{Some of the data presented herein were obtained at the W. M. Keck Observatory, which is operated as a scientific partnership among the California Institute of Technology, the University of California, and the National Aeronautics and Space Administration. The Observatory was made possible by the generous financial support of the W. M. Keck Foundation.}}

\author[T. Fetherolf et al.]{Tara Fetherolf,$^1$\thanks{E-mail: Tara.Fetherolf@gmail.com (TF)}
Naveen A. Reddy,$^1$
Alice E. Shapley,$^2$
Mariska Kriek,$^3$
\newauthor
Brian Siana,$^1$
Alison L. Coil,$^4$
Bahram Mobasher,$^1$
William R. Freeman,$^1$
\newauthor
Ryan L. Sanders,$^5$\thanks{Hubble Fellow}
Sedona H. Price,$^6$
Irene Shivaei,$^7$\samethanks\
Mojegan Azadi,$^8$
\newauthor
Laura de Groot,$^9$
Gene C.K. Leung,$^4$
and Tom O. Zick$^3$
\\
$^1$Department of Physics \& Astronomy, University of California, Riverside, 900 University Ave., Riverside, CA 92521, USA \\
$^2$Department of Physics \& Astronomy, University of California, Los Angeles, 430 Portola Plaza, Los Angeles, CA 90095, USA \\
$^3$Astronomy Department, University of California at Berkeley, Berkeley, CA 94720, USA \\
$^4$Center for Astrophysics and Space Sciences, Department of Physics, University of California, San Diego, 9500 Gilman Drive, La Jolla, CA 92093, USA \\
$^5$Department of Physics, University of California, Davis, 1 Shields Avenue, Davis, CA 95616, USA \\
$^6$Max-Planck-Institut F\"{u}r Extraterrestrische Physik, Postfach 1312, Garching, D-85741, Germany \\
$^7$Steward Observatory, University of Arizona, 933 North Cherry Avenue, Tucson, AZ 85721, USA \\
$^8$Center for Astrophysics | Harvard \& Smithsonian, 60 Garden Street, Cambridge, MA 02138, USA \\
$^9$Department of Physics, The College of Wooster, 1189 Beall Avenue, Wooster, OH 44691, USA
}

\pubyear{2020}

\begin{document}
\label{firstpage}
\pagerange{\pageref{firstpage}--\pageref{lastpage}}
\maketitle

\begin{abstract}
% 
% NOTE: Needs to be <250 words. 
We use a sample of \nsamp~star-forming galaxies at $\zmin<z<\zmax$ from the MOSFIRE Deep Evolution Field survey to demonstrate an improved Voronoi binning technique that we use to study the properties of resolved stellar populations in $z\sim2$ galaxies. Stellar population and dust maps are constructed from the high-resolution CANDELS/3D-HST multi-band imaging. Rather than constructing the layout of resolved elements (i.e., Voronoi bins) from the S/N distribution of the $H_{160}$-band alone, we introduce a modified Voronoi binning method that additionally incorporates the S/N distribution of several resolved filters. The SED-derived resolved \EBVs, stellar population ages, SFRs, and stellar masses that are inferred from the Voronoi bins constructed from multiple filters are generally consistent with the properties inferred from the integrated photometry within the uncertainties, with the exception of the inferred \EBVs\ from our $z\sim1.5$ sample due to their UV slopes being unconstrained by the resolved photometry. The results from our multi-filter Voronoi binning technique are compared to those derived from a ``traditional'' single-filter Voronoi binning approach. We find that single-filter binning produces inferred \EBVs\ that are systematically redder by 0.02\,mag on average, but could differ by up to 0.20\,mag, and could be attributed to poorly constrained resolved photometry covering the UV slope. Overall, we advocate that our methodology produces more reliable SED-derived parameters due to the best-fit resolved SEDs being better constrained at all resolved wavelengths---particularly those covering the UV slope. 
\\ \\ \\% Spacing correction
\end{abstract}

\section{Introduction} 
Quantifying the distributions of stars, gas, and dust is an important step in understanding the formation and evolution of galaxies. For example, the locations of massive stars and their rates of formation can help address the assembly history of galaxies \citep[e.g.,][]{Wuyts12, Hemmati14, Boada15}, the rate and efficiency with which gas is converted into stars \citep[e.g.,][]{Lang14, Jung17, Tacchella18}, and a galaxy's merger history \citep[e.g.,][]{Conselice03, Lotz04, Lotz08, Cibinel15}. Studying the structure of galaxies at $z\sim2$ is particularly relevant as cosmic star formation reaches its peak at this epoch, and galaxies were rapidly assembling their stellar mass \citep[see][]{Madau14}. 

Galactic structure can either be assessed qualitatively by eye (e.g., Hubble's Tuning Fork; \citealt{Hubble26}) or quantitatively using parametric (e.g., S\'{e}rsic profile; \citealt{Sersic63}) and non-parametric (e.g., CAS, Gini, $M_{20}$; \citealt{Conselice03, Abraham03, Lotz04}) metrics that measure the distribution of light. However, morphological classification schemes can potentially break down at higher redshifts. Compared to typical local galaxies, high-redshift galaxies are at an earlier stage in their evolution and appear clumpier and irregular in shape \citep[e.g.,][]{Griffiths94, Dickinson00, Papovich05, Shapley11, Conselice14}. Furthermore, since galaxies at $z\sim2$ are in the process of building the bulk of their stellar mass \citep{Madau14}, they are typically physically smaller in size due to their lower stellar masses \citep[e.g.,][]{Toft07, Trujillo07, van_Der_Wel14, Scott17, van_de_Sande18} compared to their low-redshift counterparts. Observationally, these distant galaxies are also subject to the effects of cosmological dimming \citep[see][]{Barden08}, and the highest spatial resolution achievable with space-based facilities like the {\em Hubble Space Telescope} ({\em HST}) cover physically larger regions (few-kpc per resolution element) compared to that of lower redshift galaxies that are observed with the same instruments (sub-kpc per resolution element). Consequently, careful analyses are required when quantifying the morphology of galaxies (i.e., the distribution of stars, gas, and dust within galaxies) across a range of redshifts in an effort to infer how the structure of galaxies changes with cosmic time. 

Several studies have explored resolved stellar populations in $z\gtrsim2$ galaxies by fitting the observed pixel-to-pixel spectral energy distribution (SED) with stellar population synthesis models \citep{Guo12-1, Guo15, Guo18, Boada15, Jung17, Tacchella18}. While the highest resolution (for a given pixel scale) is obtained by fitting for the resolved SEDs within individual pixels, the signal-to-noise (S/N) within individual pixels may be too low such that the derived resolved stellar population properties may be unreliable. Furthermore, the signal may be correlated between adjacent pixels given that the spatial resolution of the instrument (i.e., the point-spread function, or PSF) is larger than the size of the individual pixels. To compensate for these effects, several studies have implemented an adaptive Voronoi binning technique \citep{Cappellari03} that groups pixels based on their S/N distribution such that low S/N pixels are grouped together into larger bins \citep[e.g.,][]{Wuyts12, Wuyts13, Wuyts14, Genzel13, Tadaki14, Lang14, Chan16}. Typically, Voronoi bins are constructed using the S/N in a single filter, namely the $H_{160}$ band.  However, for redder galaxies, such construction can lead to Voronoi bins with low S/N in bluer bands, resulting in poor constrains on the resolved stellar populations.  If the S/N is not sufficiently high in at least 5 resolved filters \citep[see][]{Torrey15}, then the resulting best-fit SEDs may not produce reliable stellar population inferred quantities for these bins. 

In this work, we introduce an improved Voronoi binning method that groups pixels based on reaching a desired S/N threshold in multiple resolved filters. Our method improves the reliability of the derived stellar population parameters---albeit by sacrificing some spatial resolution---compared to the traditional single-filter approach for constructing Voronoi bins. In this paper, we apply our methodology to a sample of \nsamp~star-forming galaxies drawn from the MOSFIRE Deep Evolution Field survey \citep[MOSDEF;][]{Kriek15}. The MOSDEF survey obtained rest-frame optical spectra for $\sim$1500~star-forming galaxies and AGNs at $1.4\lesssim z \lesssim3.8$ that were drawn from the Cosmic Assembly Near-IR Deep Legacy Survey \citep[CANDELS;][]{Grogin11, Koekemoer11} and overlap with observations from the 3D-HST survey \citep{Brammer12, Skelton14}. Our study of the resolved stellar populations in high-redshift galaxies improves upon previous work.  Previous resolved studies of $z\gtrsim2$ galaxies either lacked accurate spectroscopic redshifts necessary to mitigate degeneracies in SED-fitting (typically only photometric redshifts were used) or had samples with fewer than 100~galaxies \citep{Wuyts12, Guo12-1, Guo15, Guo18, Tadaki14, Boada15, Jung17, Tacchella18}. The statistically large sample of galaxies with robust spectroscopic redshifts and emission line measurements in the MOSDEF dataset, on the other hand, enables an unprecedented view into the resolved stellar populations of $z\sim2$ galaxies. In this paper, we focus on the methodology of our Voronoi binning technique and demonstrate its utility.  Detailed analysis of the resolved stellar population parameters obtained based on the new methodology will be presented in a follow-up paper.  

% UPDATE if sections are reorganized
The data, sample selection, and integrated stellar population synthesis modeling is presented in \autoref{sec:data}. In \autoref{sec:methods}, our modified multi-filter Voronoi binning technique for building stellar population and reddening maps is outlined. The resolved SEDs and SED-derived properties are compared to unresolved global quantities derived from the 3D-HST photometry in \autoref{sec:integrated} and our modified multi-filter Voronoi binning method is compared to the traditional single-filter Voronoi binning technique. Finally, the key points are summarized and we provide recommendations for resolved SED fitting analyses in \autoref{sec:summary}.

All line wavelength measurements are given in vacuum and magnitudes are expressed in the AB system \citep{Oke83} throughout this paper. A cosmology with $H_0=70$\,km\,s$^{-1}$\,Mpc$^{-1}$, ${\Omega_{\Lambda}=0.7}$, and ${\Omega_m=0.3}$ is assumed.

\section{Data, Sample Selection, and Integrated SED-derived Properties}\label{sec:data}
\begin{figure*}
\includegraphics[width=\textwidth]{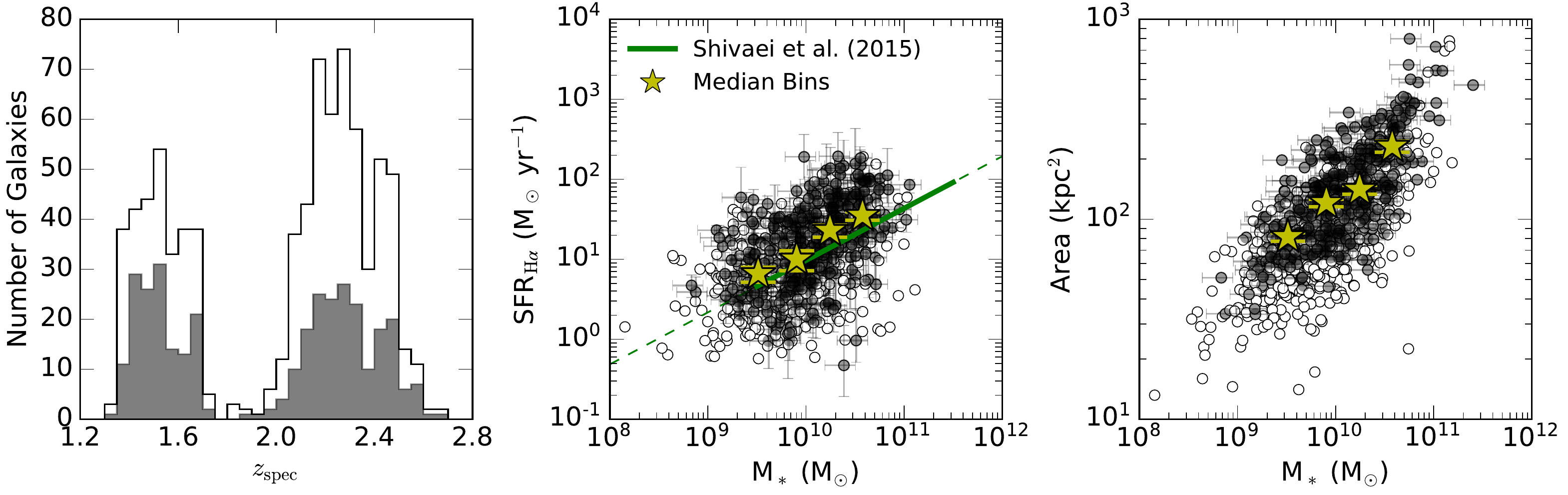}
\caption{\textit{Left:} Spectroscopic redshift distribution of galaxies in our sample covering two redshift bins: $\izmin<z<1.74$ and $1.84<z<\zmax$. The white histogram shows the initial sample distribution (\insamp~galaxies). The gray histogram denotes our primary sample (\nsamp~galaxies) that results after S/N restrictions are applied in the Voronoi binning procedure (see \autoref{sec:vbin}). \textit{Middle:} \HA\ SFR versus stellar mass, i.e. the star-forming main sequence. \HA\ emission is corrected for dust using the Balmer decrement and assuming the \citet{Cardelli89} extinction curve. All \HA\ SFRs are calculated assuming a \citet{Chabrier03} IMF and 20\% Solar metallicity (see \autoref{sec:sedfit}). The solid green line (extended by the dashed line for visual clarity) is the best linear fit to the SFR--$M_*$ relation based on the first two years of data from the MOSDEF survey \citep{Shivaei15}. The empty and filled circles show the galaxies in the initial (\insamp) and final samples (\nsamp), respectively. The yellow stars are the medians of the individual points of the \nsamp~galaxies in our sample, equally divided into four bins of stellar mass. \textit{Right:} The size--$M_*$ relation for our initial and selected samples, with the same symbol coloring as the middle panel. The area is measured from the detected 3D-HST segmentation map region for each galaxy. Our final sample selection is marginally biased against compact, low-mass galaxies due to the S/N requirements imposed by the Voronoi binning procedure (see \autoref{sec:vbin}).}
\label{fig:sample}
\end{figure*}
%
% UPDATE if subsections are reorganized
In this section, we present our observations, sample selection, and stellar population parameters inferred from fitting unresolved (integrated, or global) photometry. In \autoref{sec:3dhst}, we discuss the high-resolution, multi-wavelength CANDELS/3D-HST photometry from which we construct the resolved stellar population and reddening maps. The MOSDEF survey is described in \autoref{sec:mosdef}, including the spectroscopic redshift measurements. The sample drawn from the parent MOSDEF survey is discussed in \autoref{sec:sample}. Finally, in \autoref{sec:sedfit}, we list the assumptions used to obtain global and resolved stellar population parameters. 

\subsection{CANDELS/3D-HST Photometry}\label{sec:3dhst}
We use multi-wavelength broadband photometry from 0.3 to 8.0\,$\mu$m in the five CANDELS \citep{Grogin11, Koekemoer11} extragalactic fields: AEGIS, COSMOS, GOODS-N, GOODS-S, and UDS. CANDELS {\em HST} imaging covers $\sim$900\,arcmin$^2$ and is 90\% complete at $H_{160}\sim25$\,mag. We make use of the publicly-available\footnote{\url{https://3dhst.research.yale.edu/}} imaging and photometric catalogs that were processed and compiled by the 3D-HST grism survey team \citep{Brammer12, Skelton14, Momcheva16}. The {\em HST} images were drizzled to a 0\farcs06\,pixel$^{-1}$ scale and PSF-convolved to the same 0\farcs18 spatial resolution as the $H_{160}$ data.

\subsection{MOSDEF Spectroscopy}\label{sec:mosdef} 
Targets for the MOSDEF survey were selected from the five CANDELS extragalactic fields using the 3D-HST photometric and spectroscopic catalogs in three redshift bins ($1.37<z<1.70$, $2.09<z<2.61$, and $2.95<z<3.80$) such that the following strong rest-frame optical emission lines fall in near-IR windows of atmospheric transmission: [OII]$\lambda3727,3730$, \HB, [OIII]$\lambda\lambda4960,5008$, \HA, [NII]$\lambda\lambda6550,6585$, and [SII]$\lambda6718,6733$. The targets were selected to an $H_{160}$-band limit of 24.0, 24.5, and 25.0\,mag for the $z=1.37-1.70$, $2.09-2.61$, and $2.95-3.80$ redshift bins (hereafter referred to as the $z\sim1.5$, $z\sim2.3$, and $z\sim3$ samples), respectively, corresponding to a stellar mass limit of $\sim$10$^9$\,\Msun~for all three redshift bins. The galaxy identifications used throughout this paper refer to the 3D-HST v4 photometric catalog.

The 4.5\,year MOSDEF survey obtained rest-frame optical spectra for $\sim$1500 star-forming and AGN galaxies using the MOSFIRE multi-object spectrograph \citep{McLean10, McLean12} in the $Y$, $J$, $H$, and $K$ bands ($R=$ 3400, 3000, 3650, and 3600 using 0\farcs7 slit widths) on the 10\,m Keck I telescope. Every slit mask also included at least one slit star, which is used for the absolute flux calibration and corrections for slit loss. Line fluxes are measured by fitting a linear function to the continuum and a Gaussian function to each emission line. For the [OII]$\lambda\lambda3727,3730$ doublet and \HA+[NII]$\lambda\lambda6550,6585$ lines, a double and triple Gaussian is fit the the lines, respectively. The stellar population model that best fits the observed 3D-HST photometry (\autoref{sec:sedfit}) is used to correct \HA\ and \HB\ line fluxes for underlying Balmer absorption. Flux errors are obtained from the 68th-percentile width of the distribution of line fluxes that have been remeasured from the 1D spectra that have been perturbed by their error spectra 1000 times. Finally, the spectroscopic redshift is measured from the observed wavelength of the highest signal-to-noise (S/N) emission line, which is typically \HA\ or [OIII]$\lambda5008$. Refer to \citet{Kriek15} and \citet{Reddy15} for additional details on the MOSDEF survey, including the observing strategy, data reduction, and line flux measurements. 

In this study, the redshift of each galaxy (and its resolved components) is fixed to the measured MOSDEF spectroscopic redshift in order to constrain the stellar population parameters derived from the SED fitting. 

\subsection{Sample Selection}\label{sec:sample}
Our sample is derived from the MOSDEF parent sample in the $z\sim1.5$ and $z\sim2.3$ redshift bins. First, only galaxies with robust spectroscopic redshifts that have been measured using at least two emission features with a S/N $>2$ are selected. AGNs are removed based on their identification through their X-ray luminosities, optical emission lines ($\log{(\text{[NII]/H}\alpha)} > -0.3$), and/or mid-IR luminosities \citep{Coil15, Azadi17, Azadi18, Leung19}. This initial sample contains \insamp~star-forming galaxies at $\izmin<z<\zmax$, including 74~galaxies that were serendipitously detected outside of the targeted MOSDEF redshift bins. Additional constraints to the S/N and resolution of the photometry are applied when performing adaptive Voronoi binning (see \autoref{sec:vbin}), which brings our sample to \nsamp~star-forming galaxies at redshifts $\zmin<z<\zmax$.

The spectroscopic redshift distribution, SFR--$M_*$ relation, and size--$M_*$ relation are shown in \autoref{fig:sample} for the galaxies in our initial (white histogram; empty circles) and final (gray histogram; filled circles) samples. Also shown is the median SFR, size, and stellar mass of our sample divided into four equally sized bins (yellow stars). The SFR bins are on average 0.16\,dex above the main sequence relation defined in \citet{Shivaei15}, which is within the intrinsic scatter of the relation. Our final selected sample is marginally biased against low-mass and low-SFR galaxies, primarily due to the tendency of such galaxies to be compact and/or faint. It is difficult to spatially resolve compact galaxies into several components and faint galaxies do not have sufficient S/N for measuring reliable spatially resolved fluxes across several filters. Therefore, these factors cause low-mass and low-SFR galaxies to be excluded from our sample during the Voronoi binning procedure (see \autoref{sec:vbin}).

\subsection{Spatially-Unresolved SED Fitting}\label{sec:sedfit}
The stellar populations are modeled on both resolved (see \autoref{sec:sedfit_res}) and unresolved scales for galaxies in our sample to obtain SED-derived \EBVs, stellar population ages, SFRs, and stellar masses. Here we describe the SED fits to the integrated galaxy light, as measured from the 3D-HST photometric catalogs. A $\chi^2$ minimization technique is used to select the \citet{Bruzual03} stellar population model that best fits the observed photometry \citep[see][]{Reddy12-1}. First, the photometry is corrected for the strongest emission lines measured in the MOSDEF spectroscopy, including \HA, \HB, [OIII]$\lambda\lambda4960,5008$, and the [OII]$\lambda3728$ doublet. In the SED modeling, we assume a \citet{Chabrier03} initial mass function (IMF), constant star-formation histories (SFHs), and allow the ages to range between 50\,Myr and the age of the Universe at the redshift of each galaxy.\footnote{\citet{Reddy12-1} found that both exponentially rising and constant SFHs best reproduce SFRs for $z\sim2$ galaxies when stellar population ages are limited to being older than the typical dynamical timescale (50\,Myr). Stellar population ages that are derived using constant SFHs are typically $\sim$30\%~younger than those assumed with exponentially rising SFHs \citep{Reddy12-1}.} Reddening in the range $0.0 \le \EBVs\ \le 0.4$ is considered while assuming an SMC extinction curve \citep{Fitzpatrick90, Gordon03} and sub-solar metallicity (0.2\,$Z_{\odot}$).\footnote{Recent studies have found that SMC-like or steeper attenuation curves paired with sub-Solar metallicities are more appropriate for young, high-redshift galaxies \citep{Reddy18-1}. The \citet{Calzetti00} attenuation curve paired with Solar metallicities may be more appropriate for massive galaxies, but primarily affects the absolute mass measurements and not their relative order \citep{Reddy18-1} and, thus, does not significantly change our results.\label{fn:atten_curve}} 

Typical SED parameter errors are derived using a subset of 50 randomly selected galaxies from our sample that span the SED fitting parameter space in \EBVs, stellar population age, SFR, and stellar mass in both redshift bins. Only 50 galaxies are selected for the error analysis in the interest of limiting computation time, but the range of SED parameters probed by these 50 galaxies is similar to that of the larger sample. The measured fluxes are perturbed 100 times by their flux errors and refit. The width of the parameter space of the 68 (of 100) models with the lowest $\chi^2$ relative to the observed photometry are assumed to represent the errors in the SED parameters. The typical SED parameter error is then estimated as the average of the errors from all galaxies in the subsample. The typical random SED parameter errors derived from the SEDs fit to the perturbed 3D-HST photometry are as follows: 0.01 in \EBVs, 0.20\,dex in log stellar population age, 0.05\,dex in log SFR, and 0.16\,dex in log stellar mass. 

\section{Resolved Stellar Population and Reddening Maps}\label{sec:methods}
% NO CONTENT HERE

\subsection{Resolved Photometry}\label{sec:photometry}
The resolved stellar population and reddening maps are constructed using \HST/WFC3 resolved imaging in the F125W, F140W, and F160W filters (hereafter $J_{125}$, $JH_{140}$, and $H_{160}$), and \HST/ACS resolved imaging in the F435W, F606W, F775W, F814W, and F850LP filters (hereafter $B_{435}$, $V_{606}$, $i_{775}$, $I_{814}$, and $z_{850}$). Note that not all of the filters are available for all of the CANDELS fields. The resolved imaging is supplemented with unresolved \textit{Spitzer}/IRAC photometry at 3.6\,$\mu$m, 4.5\,$\mu$m, 5.8\,$\mu$m, and 8.0\,$\mu$m, which has been corrected for contamination by neighboring sources using models of the \HST\ images that have been PSF-smoothed to the lower resolution IRAC photometry. Pixels associated with the 3D-HST photometry are identified using the Source Extractor \citep{Bertin96} segmentation maps provided by the 3D-HST grism survey \citep{Skelton14}, which are based on a noise-equalized combination of {$J_{125}$+$JH_{140}$+$H_{160}$}. 

RMS error (noise) maps are additionally required to measure flux uncertainties and the pixel-to-pixel S/N, which are obtained using the provided 3D-HST weight maps. While the science images have been spatially smoothed to the resolution of the $H_{160}$ filter by the 3D-HST team, the weight maps are based on the un-convolved imaging. Using the un-convolved and convolved science images, the weights are converted to spatially matched measurement errors as follows. First the weight maps are converted to noise maps using $N_{\text{orig}} = 1/\sqrt{w_{\text{orig}}}$, where $w_{\text{orig}}$ is the provided weight and $N_{\text{orig}}$ is the noise. Then the spatially matched measurement errors are calculated in each pixel by
\begin{equation}
\label{eq:noise}
N_{\text{conv}} = \left|\frac{S_{\text{conv}}}{S_{\text{orig}}/N_{\text{orig}}}\right|
\text{,}
\end{equation}
where $S_{\text{orig}}$ and $N_{\text{orig}}$ are the signal and noise of the original science images and $S_{\text{conv}}$ and $N_{\text{conv}}$ are the signal and noise of the convolved images. By definition, \autoref{eq:noise} conserves the S/N on a pixel-to-pixel level between the original and convolved images. Total flux errors measured from the convolved noise maps are typically $\sim$70\% larger than the 3D-HST photometric catalog flux errors, but the larger errors do not significantly affect our results.

\subsection{Adaptive Voronoi Binning}\label{sec:vbin}
\begin{figure*}
\includegraphics[width=.8\textwidth]{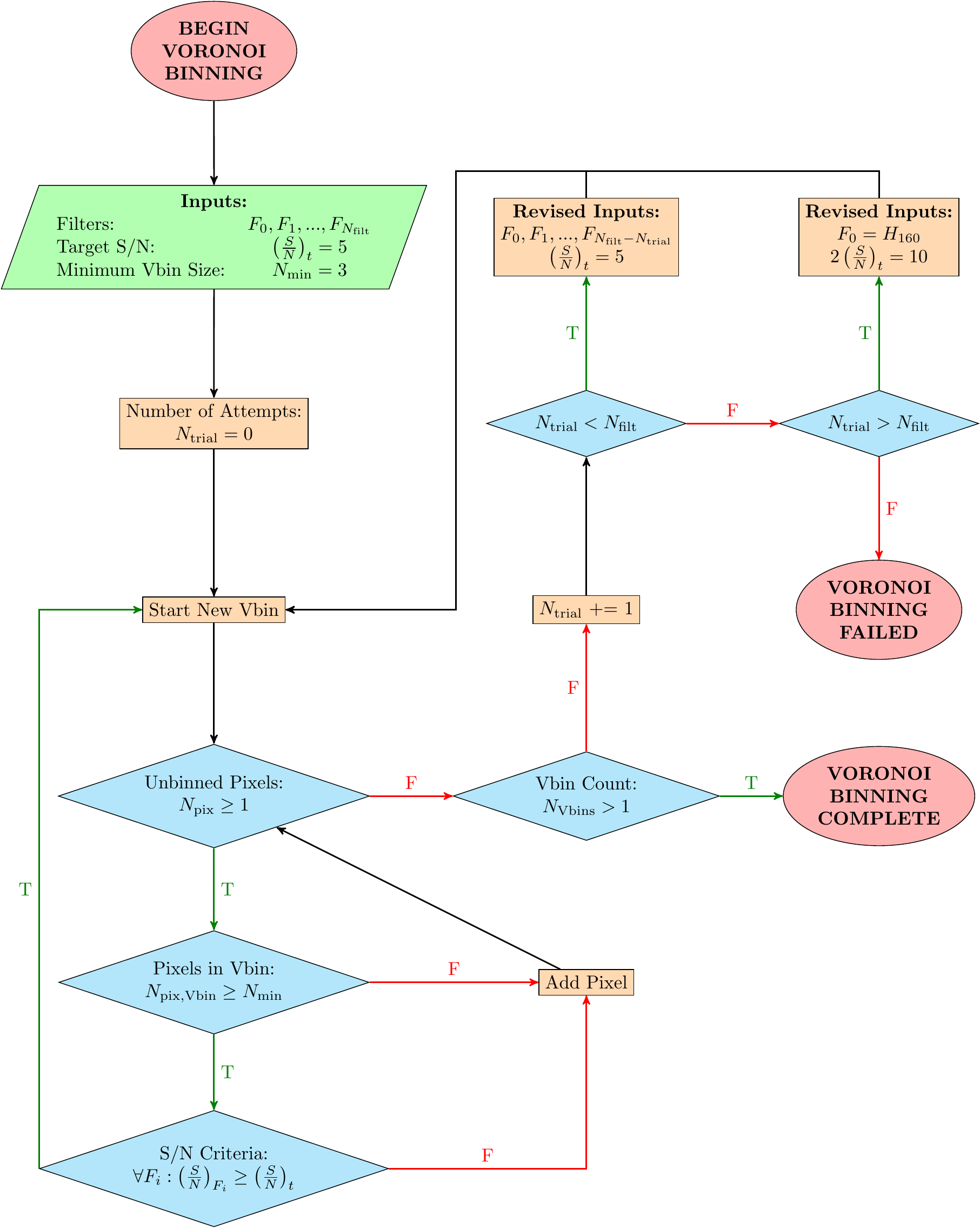}
\caption{Flowchart highlighting the modifications we applied to the \citet{Cappellari03} adaptive Voronoi binning algorithm. For this study, the primary filter is $F_0=H_{160}$, the target signal-to-noise is $(S/N)_t=5$, and the minimum number of pixels per Voronoi bin is $N_{\text{min}}=3$. $N_{\mathrm{filt}}$ is the number of additional filters used in addition to the primary filter, $N_{\mathrm{trial}}$ is the number of times adaptive Voronoi binning has been attempted, $N_{\text{pix}}$ is the number of unbinned pixels remaining, $N_{\mathrm{pix,Vbin}}$ is the number of pixels within the Voroni bin being constructed, and $N_{\text{Vbins}}$ is the total number of Voronoi bins after all pixels have been assigned into a Voronoi bin.}
\label{fig:vbin_flow}
\end{figure*}
Construction of the resolved stellar population and reddening maps begins with isolating individual galaxies in our sample. Sub-images that are $80\times80$ pixels in size (4\farcs8$\times$4\farcs8, or approximately $40\times40$\,kpc at $z\sim2$) are cut for each galaxy from the CANDELS/3D-HST imaging and its respective spatially matched noise map. The 3D-HST segmentation map is then used to mask any pixels not associated with the galaxy. The median of the pixels that are unassociated with any detected sources is used to apply a local background subtraction to each pixel of the sub-image, although the correction is negligible since the CANDELS images are already background subtracted.

To study resolved stellar populations of galaxies, SED fitting can be performed either on a pixel-by-pixel basis, or on larger ``resolved'' elements (e.g., Voronoi bins) formed by grouping together adjacent pixels. We argue that the latter procedure is necessary for a proper interpretation of the resolved stellar populations for the following reasons. Generally only the central regions of each galaxy have sufficient S/N for reliable SED fitting on a pixel-by-pixel basis. Moreover, the signal is correlated between neighboring individual pixels due to the spatial resolution indicated by the \HST\ PSF and data resampling when constructing the images.  Several studies have used the two-dimensional adaptive Voronoi binning technique developed by \citet{Cappellari03} to study resolved stellar populations in galaxies \citep[e.g.,][]{Wuyts12, Wuyts13, Wuyts14, Genzel13, Tadaki14, Lang14, Chan16}. The \citet{Cappellari03} algorithm adaptively bins pixels based on the S/N distribution such that each Voronoi bin has at least the specified target S/N while maintaining a ``roundness'' threshold. With the exception of \citet{Chan16}, these studies applied Voronoi binning to their samples by requiring a S/N~=~10 per Voronoi bin in the $H_{160}$ filter. While this approach boosts the S/N for $H_{160}$ in every resolved element, bluer filters may have lower S/N per pixel depending on the colors of the galaxy. For this reason, \citet{Chan16} instead applied the same S/N~=~10 threshold to the $z_{850}$ filter. We alternatively include the contribution of light from spatially segregated regions that are brighter at shorter wavelengths (e.g., blue star-forming regions compared to red bulge-like features) in each spatially resolved element by modifying the \citet{Cappellari03} algorithm (version 3.0.4) to construct the Voronoi bins based on the S/N distribution of multiple filters. Each filter is evaluated and is only used for Voronoi binning if the total S/N in that filter, 
\begin{equation}
\label{eq:sntot}
\left(\frac{S}{N}\right)_{\text{total}} = \frac{\sum_{i=1}^{n_{\text{pix}}} S_i}{\sqrt{\sum_{i=1}^{n_{\text{pix}}} N_i^2}}
\text{,}
\end{equation}
measured across all pixels associated with the galaxy is more than the user-designated target S/N, where $n_{\text{pix}}$ is the number of pixels identified in the segmentation map. We use a target S/N~=~5, which is acceptably lower than previous studies due to our multi-filter requirement, but necessary to obtain robust constraints on the stellar population parameters. 

The modifications applied to the \citet{Cappellari03} algorithm add constraints only while building the Voronoi bins. Unmodified processes include the pixel selection while building candidate bins, the ``roundness'' threshold of a candidate bin, or the assignment of unsuccessfully binned pixels after the bins have been created. We refer the reader to \citet{Cappellari03} for further details on their adaptive Voronoi binning technique. Here we describe how Voronoi bins are constructed given our modifications (refer to  \autoref{fig:vbin_flow} for a visual outline). The layout of the Voronoi bins is based on the S/N distribution of a primary filter ($H_{160}$ is used, similar to other studies), although the S/N distribution in all applicable filters is considered. A candidate bin begins with the unbinned pixel that has the highest S/N in the primary filter, then neighboring pixels are added to the bin until the specified target S/N is reached in all applicable filters. The modified algorithm also includes designating an optional minimum Voronoi bin size. The smallest Voroni bins are constrained to contain at least 3 pixels (in contrast with typical Voronoi binning schemes that allow single pixels to be Voronoi bins) such that they match the \HST\ PSF resolution of 0\farcs18, translating to the smallest resolution elements having an area of $\sim$0.75\,kpc$^2$. If this binning procedure fails (for example, the entire galaxy has been grouped into a single bin), the process is repeated using one less filter---specifically the filter with the lowest $(S/N)_{\text{total}}$ as calculated in \autoref{eq:sntot} (note that the primary filter is always required). Upon each failed binning attempt, this procedure will repeat iteratively until only the primary filter remains. If all other filters have been rejected, the revised adaptive Voronoi binning algorithm is applied to the primary filter ($H_{160}$) alone using a S/N that is twice that of the original target S/N (S/N~$=10$, matching the requirements used by previous studies). However, for this study we require that the bins be defined by multiple filters that span the Balmer and 4000\,\AA~breaks in order to more robustly quantify ages and stellar masses on a bin-by-bin basis. Therefore, any galaxies that failed to be adaptively binned using at least two filters ($H_{160}$ and one additional filter) are removed from our sample.

\begin{figure}
\includegraphics[width=\columnwidth]{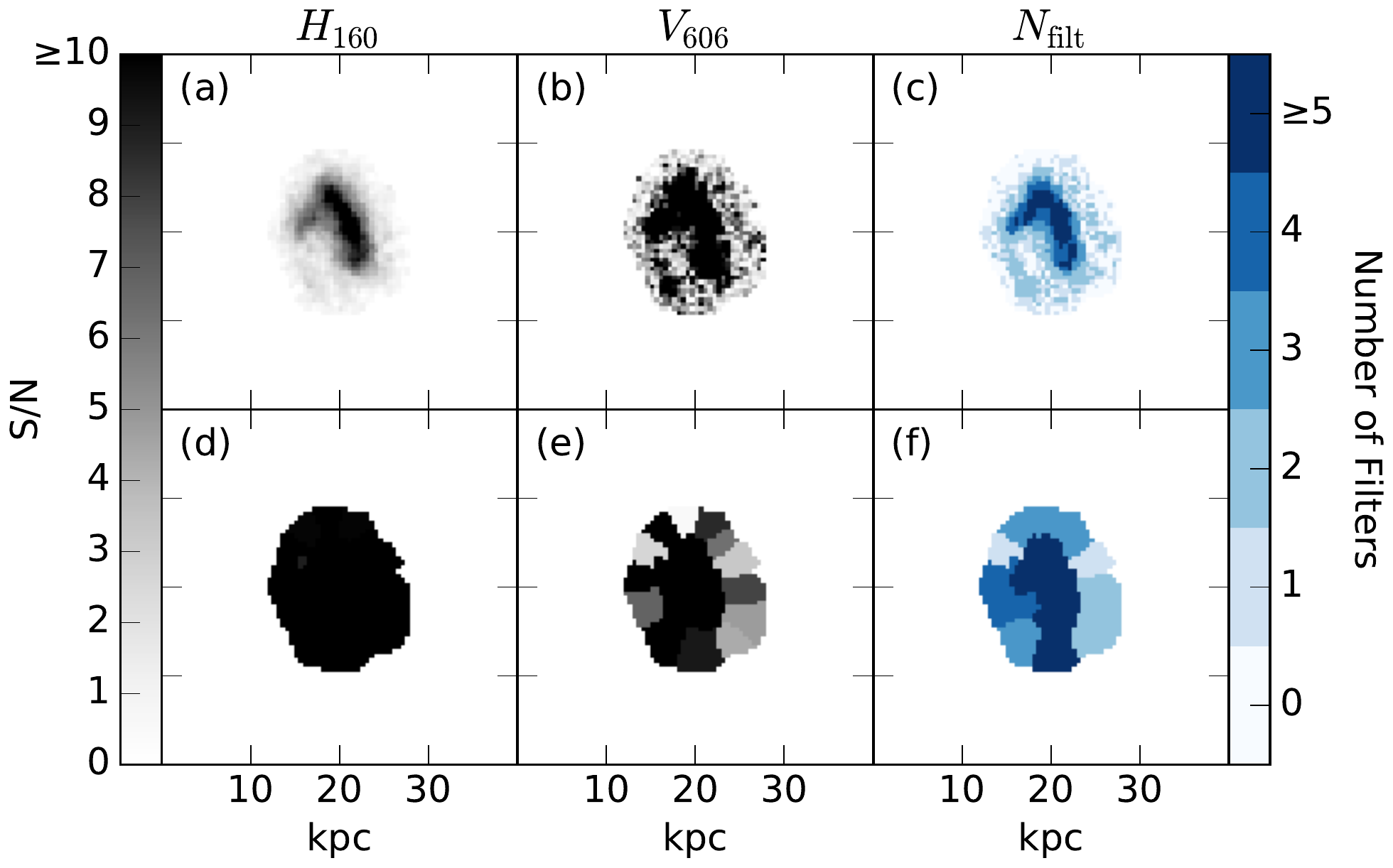}
\caption{The S/N maps before and after applying adaptive Voronoi binning for one of the galaxies in our sample, UDS\,21834 ($z=1.67$). Only the 911\,pixels and 41\,bins identified by the segmentation map are shown. \textit{Top row}: the pixel-to-pixel S/N distribution in (a) $H_{160}$ and (b) $V_{606}$. Panel (c) shows for each pixel the number of filters ($N_{\text{filt}}$) that have a S/N~$\geq5$. \textit{Bottom row}: the Voronoi bin S/N distribution in the (d) $H_{160}$ and (e) $V_{606}$ filters. Panel (f) shows for each Voronoi bin the number of filters that have a S/N~$\geq5$. The Voronoi bins within the darkest blue region match our criteria of a S/N~$\geq5$ in $\geq$5 filters, which throughout the text will be referred to simply as the ``Voronoi bins.'' All other lighter blue regions make up the ``outskirt'' component and will be referred to as such throughout the text.}
\label{fig:vbin_steps}
\end{figure}
As an example, \autoref{fig:vbin_steps} shows how the S/N distribution for one of the galaxies in our sample (UDS\,21834, $z=1.67$) varies between the $H_{160}$ and $V_{606}$ filters on the pixel-to-pixel level in panels (a) and (b) and across the Voronoi bins in panels (d) and (e). Panel (f) in \autoref{fig:vbin_steps} shows the number of filters that have a S/N~$\geq5$ in each Voronoi bin. For comparison, panel (c) shows the number of filters that have a S/N~$\geq5$ on the pixel-to-pixel level. \citet{Torrey15} found that applying a simplified set of SED fitting assumptions (i.e., dust-free, restricted ages, and limited metallicities) to a sample of simulated galaxies observed in 5 filters reproduced stellar masses within a factor of two of the known simulated mass. Therefore, reliable best-fit SEDs are expected to be produced from the regions where a S/N~$\geq5$ is reached in at least 5 filters, as is represented by the darkest blue color in panels (c) and (f) of \autoref{fig:vbin_steps}. By comparing the areas covered by the darkest blue color in panels (c) and (f), it can be seen that a larger region of the galaxy can be used for reliable SED fitting when adaptive Voronoi binning is applied with a multi-filter requirement. For this reason, we only consider Voronoi bins that have a S/N~$\geq5$ in a minimum of 5 filters for the primary SED fitting. Throughout the text, we refer to the bins with a S/N~$\geq5$ in at least 5 filters as the ``Voronoi bins.'' We also ensure that the high S/N filters cover both sides of the Balmer and 4000\,\AA~break region, which further constrains the stellar population age estimates. Bins that are not above the target S/N contain the ``remaining'' signal and tend to be in the outskirts of the galaxy (light blue bins shown in panel (f) of \autoref{fig:vbin_steps}). The bins that have less than 5 filters with S/N~$\geq5$ are referred to as ``outskirt'' components of the galaxy. Collectively, the high S/N Voronoi bins combined with the outskirt bins cover exactly the same area defined by the 3D-HST segmentation map. In \autoref{sec:integrated}, the integrated SED and its respective parameters derived from the conservatively selected Voronoi bins are compared to those derived from additionally including the outskirt components. Outskirt bins are only included in our analyses where specifically noted as their SED-derived properties are more uncertain due to the lower S/N in these bins.

\begin{figure}
\includegraphics[width=\columnwidth]{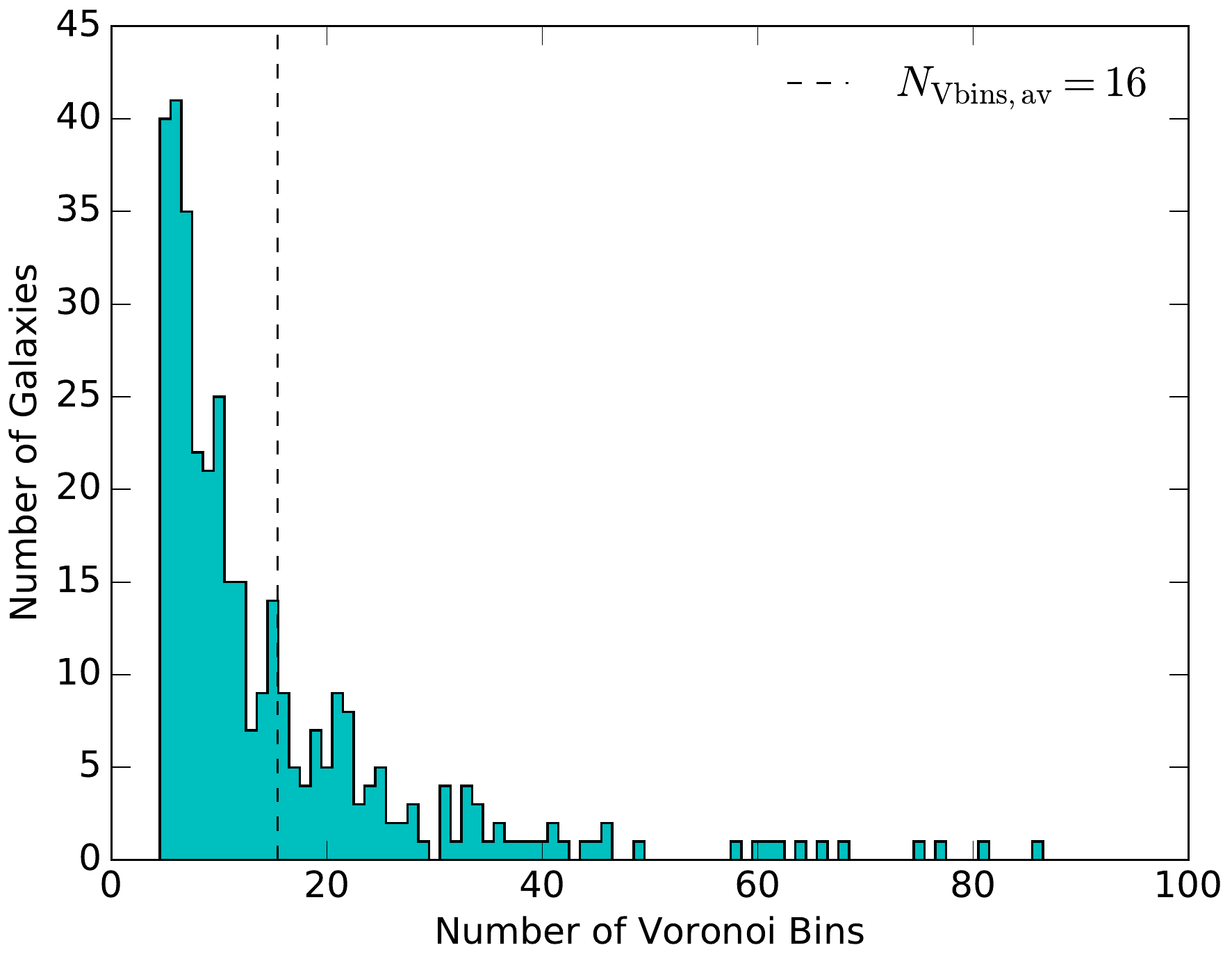}
\caption{Distribution of the number of Voronoi bins within each galaxy in our sample. To be included in our sample, galaxies must have at least 5 Voronoi bins. The black dashed vertical line shows the average number of Voronoi bins per galaxy, $N=16$. The median Voronoi bin size is 6\,pixels and the minimum Voronoi bin size is set to 3\,pixels.}
\label{fig:nVbins}
\end{figure}

Galaxies that have less than 5 Voronoi bins (outskirt bins do not contribute) remaining are removed from our sample, as they are considered unresolved. Galaxies in our sample have 5--151 Voronoi bins ($\sim$16\,bins on average), as is shown by the histogram in \autoref{fig:nVbins}. On average, an individual Voronoi bin contains $\sim$18\,pixels ($\sim$4.5\,kpc$^2$) and the median bin size is 6\,pixels (1.5\,kpc$^2$). Overall the Voronoi binning procedure reduces the sample size from the \insamp~star-forming galaxies discussed in \autoref{sec:sample} to \nsamp~galaxies. As can be seen in \autoref{fig:sample}, the Voronoi binning requirements marginally biases our sample against low-mass, low-SFR, and compact galaxies. 

\subsection{Flux Measurements}\label{sec:flux}
The counts in the convolved science images and noise maps are converted to flux density, and the AB magnitude and magnitude error are calculated within each Voronoi bin. The minimum magnitude error is set to 0.05\,mag, in order to prevent any single photometric point from skewing the SED fit. 

To better constrain the stellar masses estimated from the SED fitting, \textit{Spitzer}/IRAC unresolved photometry (normalized to the segmentation map area) is paired with the resolved photometry. We choose to simply normalize the IRAC photometry by the $H_{160}$ flux in each resolved element, rather than adjusting the best-fit resolved SEDs to match the integrated 3D-HST SED as suggested by \citet{Wuyts12}. The IRAC flux in each resolved element, $F_{\text{IRAC,Vbin}}$, is estimated by normalizing the total IRAC flux by the $H_{160}$ bin flux. The normalized IRAC flux per Voronoi bin is defined as
\begin{equation}
\label{eq:irac}
F_{\text{IRAC,Vbin}}=F_{\text{IRAC,tot}}\frac{H_{160,\text{Vbin}}}{H_{160,\text{tot}}}
\text{,}
\end{equation}
where $H_{160,\text{tot}}$ and $F_{\text{IRAC,tot}}$ are the 3D-HST $H_{160}$ and IRAC broadband fluxes, and $H_{160,\text{Vbin}}$ is the $H_{160}$ flux measured within the resolved element (Voronoi bin, in our case). In \aref{app:wuyts}, we show that the differences in the best-fit SEDs and SED-derived parameters are typically insignificant when using the more computationally-expensive methodology described by \citet{Wuyts12}.  However, the \citet{Wuyts12} method may be more appropriate when key SED features are not properly constrained by the resolved photometry (e.g., UV slope, Balmer/4000\,\AA\ breaks; also see \autoref{sec:intparam}).  

Using the normalization defined by \autoref{eq:irac} forces the \HIRAC\ colors to be constant across the entire galaxy, thus constraining the strength of the Balmer and 4000\,\AA~breaks for the highest redshift galaxies in our sample ($z\gtrsim2.5$). As the Balmer/4000\,\AA~break is sensitive to stellar population age and attenuation \citep[e.g.,][]{Worthey94, Shapley01}, we must be wary of how fixing the \HIRAC\ color affects the derived age and reddening maps. In \aref{app:colors}, we investigate how variations in the \HIRAC\ colors influence the distribution of stellar population ages and reddening inferred through the SED fitting. In general, we find that the redder Voronoi bins remain red (higher \EBVs) and the bluer Voronoi bins remain blue (lower \EBVs) when their \HIRAC\ colors are perturbed, indicating that our constraint on the \HIRAC\ colors (i.e., \autoref{eq:irac}) does not unduly influence the results concerning the distribution of stellar population ages and attenuation within galaxies.

\subsection{Resolved SED Fitting}\label{sec:sedfit_res}
\begin{figure*}
\begin{adjustbox}{width=\linewidth, center}
\begin{minipage}{\linewidth}
\includegraphics[width=\linewidth]{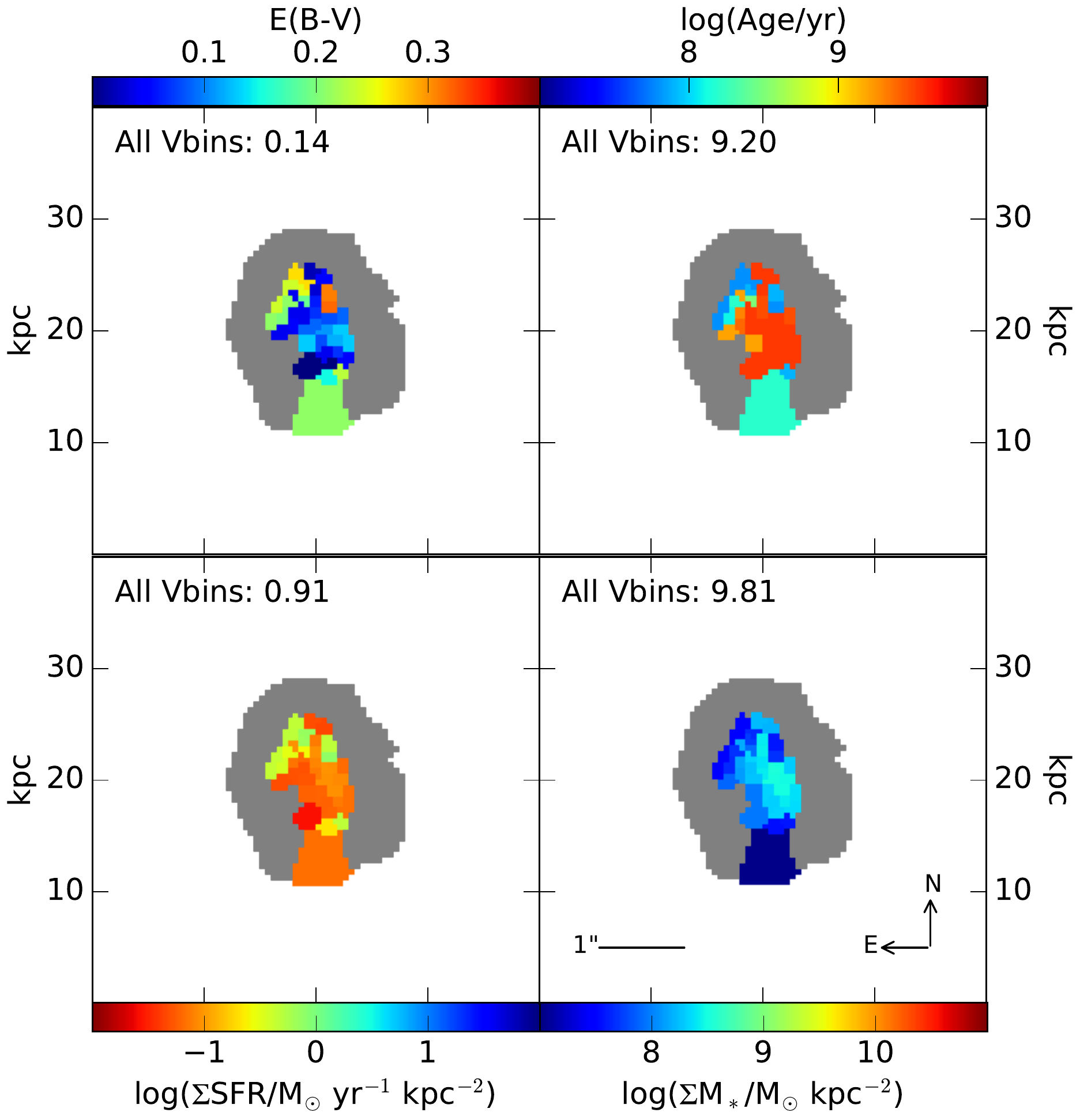}
\end{minipage}
\quad
\begin{minipage}{\linewidth}
\includegraphics[width=\linewidth]{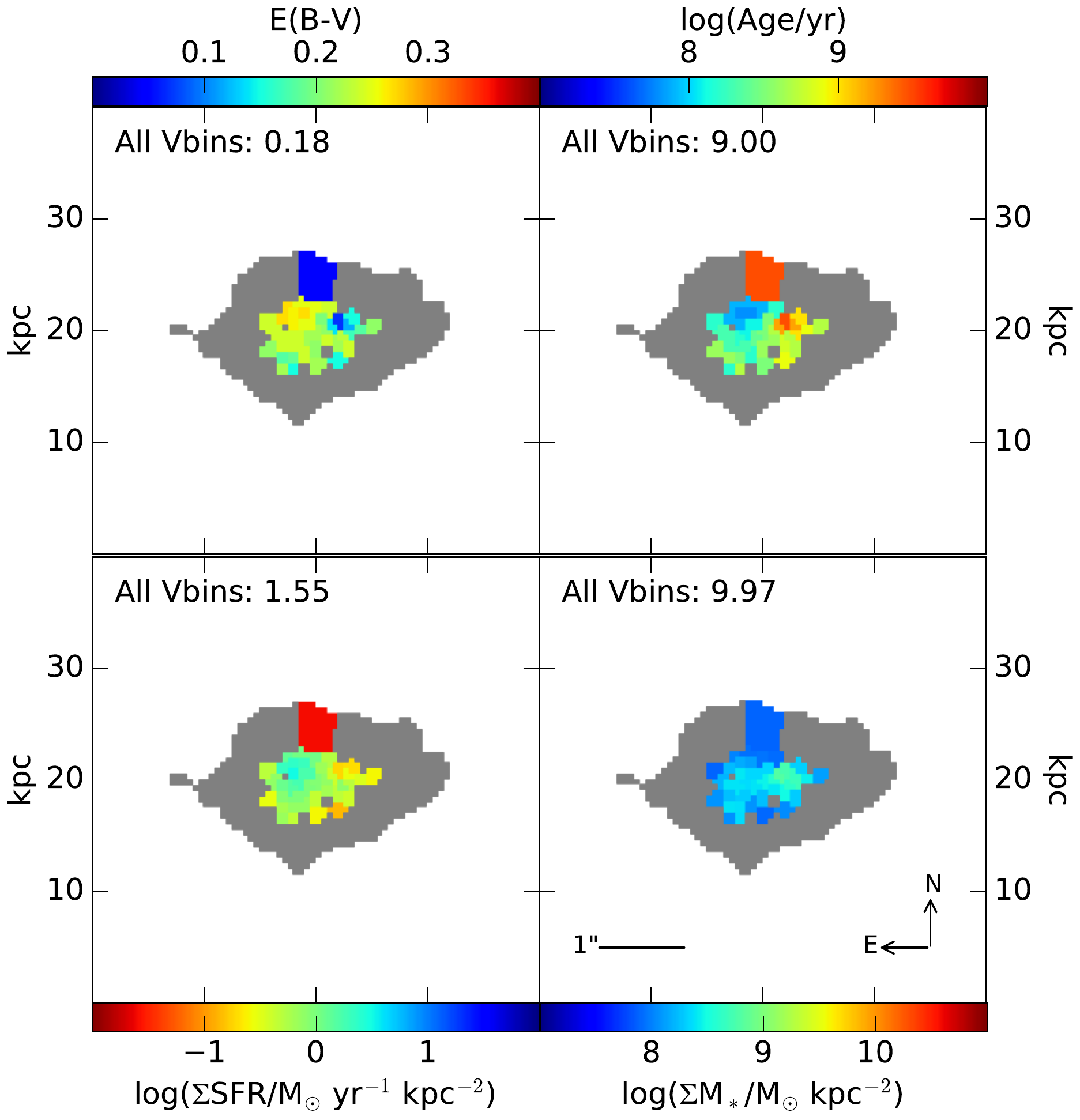}
\end{minipage}
\end{adjustbox}
\caption{Stellar population and reddening maps from resolved SED fitting for UDS\,21834 ($z=1.67$, \textit{left panels}) and COSMOS\,3666 ($z=2.09$, \textit{right panels}), showing the distribution of \EBVs\ (\textit{top left}), stellar population age (\textit{top right}), SFR (\textit{bottom left}), and stellar mass (\textit{bottom right}). These galaxies both have 41\,Voronoi bins with the bins containing an average of 8 and 6\,pixels ($\sim$2\,kpc$^2$ and $\sim$1.5\,kpc$^2$), respectively. The gray regions indicate the outskirt component consisting of regions that are included in the segmentation map, but do not fulfill the Voronoi bin criteria (S/N $\geq5$ in at least 5 filters) discussed in \autoref{sec:vbin}. The listed number in each panel shows the average \EBVs\ or stellar population age weighted by the Voronoi bin areas or the summed SFR or stellar mass from all of the Voronoi bins.}
\label{fig:map_ex}
\end{figure*}
\begin{figure*}
\includegraphics[width=.9\textwidth]{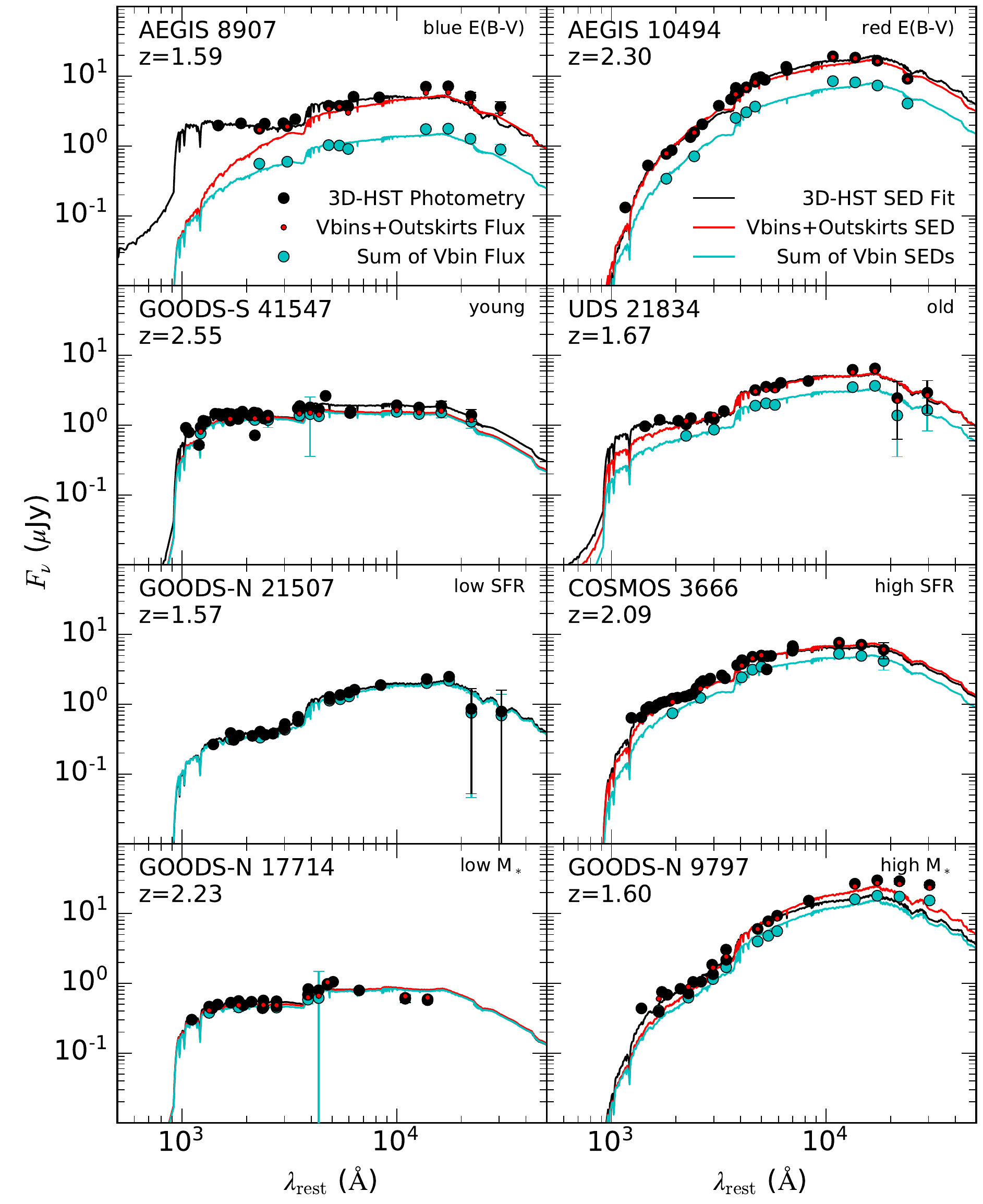}
\caption{Best-fit SEDs for eight example galaxies spanning the SED fitting parameter space. From top to bottom, each row respectively shows example galaxies with low (\textit{left}) and high (\textit{right}) \EBVs, stellar population ages, SFRs, and stellar masses. The black points are the fluxes from the 3D-HST photometric catalog \citep{Skelton14} corrected for the brightest MOSDEF emission lines and the black curves are their respective best-fit SEDs. The light blue points are the total integrated fluxes within the Voronoi bins (Voronoi bins with S/N~$\geq5$ in at least 5 different filters). The light blue curves represent the sum of the SEDs that are best fit to the individual Voronoi bins.  The red points indicate the total flux obtained when summing the Voronoi bin and outskirt bin fluxes, which naturally lie close to the values obtained from the 3D-HST photometry. The red curves show the SED fits to the red points.  Photometric errors are typically smaller than the size of the points. Note that not all galaxies have outskirt components, such as GOODS-N\,21507.}
\label{fig:sedcomp}
\end{figure*}
The same model assumptions that were outlined in \autoref{sec:sedfit} are also used for the resolved SED fitting. For each galaxy in our sample, two representations of the integrated properties are obtained by fitting a single SED to (a) the multi-wavelength 3D-HST broadband photometry \citep[][see \autoref{sec:sedfit}]{Skelton14} and (b) the total sum of the flux within all of the valid Voronoi bins (including the total 3D-HST IRAC fluxes). Resolved SED modeling is performed on all of the individual Voronoi bins and outskirt components. In \autoref{sec:integrated} we compare the resolved and integrated SED fitting results and demonstrate how using only the high S/N Voronoi bins does not significantly affect the results from the SED fitting (see \autoref{sec:vbin}). \autoref{fig:map_ex} shows the resolved stellar population and reddening maps for UDS\,21834 ($z=1.67$) and COSMOS\,3666 ($z=2.09$) as examples. The average \EBVs, average stellar population age, summed SFR, and summed stellar mass derived from the Voronoi bins is listed in the top left corner of each panel in \autoref{fig:map_ex}.

Typical resolved SED parameter errors are derived using the same subset of 50~galaxies as was used to obtain the integrated SED parameter errors (see \autoref{sec:sedfit}). Only 50~galaxies are selected for the error analysis in the interest of limiting computation time, but the range of SED parameters probed by these 50 galaxies is similar to that of the larger sample. The measured fluxes in each Voronoi bin are perturbed 100 times by their flux errors and refit. The width of the distribution of parameters for the 68 (of 100) models with the lowest $\chi^2$ relative to the observed photometry are assumed to represent the errors in those parameters for each individual Voronoi bin. Within a single galaxy, the SED parameter errors of the individual Voronoi bins are averaged together, resulting in a typical Voronoi bin error within the individual galaxy. Finally, the typical Voronoi bin SED parameter error is estimated as the average of the typical Voronoi bin errors from all 50 galaxies in the subsample. The typical SED parameter errors derived from the SEDs fit to the perturbed Voronoi bin fluxes are as follows: 0.03 in \EBVs, 0.21\,dex in log stellar population age, 0.16\,dex in log SFR, and 0.10\,dex in log stellar mass. 

\section{Resolved vs. Integrated Properties}\label{sec:integrated}
In comparing the resolved and integrated stellar population parameters, we must be aware of the different apertures over which the measurements are made, and the slightly different methodology used to fit the SEDs. Specifically, the integrated 3D-HST flux measurements are based on the flux within 0\farcs7 apertures and are corrected for the contribution from the strongest emission lines that were detected in the MOSFIRE spectra, including [OII]$\lambda3727,3730$, \HB, [OIII]$\lambda\lambda4960,5008$, and \HA\ \citep{Reddy15}. The resolved Voronoi bin fluxes, on the other hand, are measured directly from the sum of the flux detected by the pixels in the resolved imaging and are not corrected for the contribution of the emission lines detected by MOSDEF. Additionally, the integrated 3D-HST photometry covers a larger dynamic range in wavelength than the range covered by the filters that are available for resolved photometry due to the ancillary ground-based data that is available. While the \textit{Spitzer}/IRAC photometry is included to increase wavelength coverage of the resolved photometry, we assume that the \HIRAC\ colors are fixed across all resolved elements using \autoref{eq:irac}. Despite these inherent differences between the integrated and resolved photometry, the $V_{606}-H_{160}$ colors measured from the integrated 3D-HST fluxes and the sum of the flux from all Voronoi bins and outskirt components agree within the measured scatter of 0.3\,mag. 

% UPDATE if subsections are reorganized
The biases that may arise from differences between the resolved and unresolved photometry are explored by directly comparing the associated best-fit SEDs (\autoref{sec:intsed}) and SED-derived parameters (\autoref{sec:intparam}). In \autoref{sec:traditional}, we show how our modified multi-filter Voronoi binning technique can improve upon that using the $H_{160}$ filter alone by comparing the best-fit SEDs and SED-derived parameters from the two techniques. Finally, the implications of our findings on resolved stellar population studies is discussed in \autoref{sec:discussion}.

\subsection{Comparing Best-fit SEDs}\label{sec:intsed}
The unresolved and summed resolved SEDs for eight galaxies in our sample that span the SED fitting parameter space (low/high \EBVs, stellar population age, SFR, and stellar mass) are shown in \autoref{fig:sedcomp}. The summed best-fit SEDs derived from the resolved Voronoi bin photometry (light blue points and curves) are compared to the SEDs derived from the unresolved photometry (black points and curves). The 3D-HST photometry (black points) has extended wavelength coverage and includes more light than the valid Voronoi bins (light blue points)---typically 0.85\,mag brighter. The summed Voronoi bin SEDs do not include ``outskirt'' components, which are defined as low S/N regions in \autoref{sec:vbin}. Therefore, \autoref{fig:sedcomp} additionally shows the SEDs that result from adding the outskirt component SEDs to the summed Voronoi bin SEDs (red points and curves). The total flux from the sum of the Voronoi bins and outskirt components is nearly equivalent to the measured flux from the unresolved 3D-HST photometry for filters with resolved imaging (red and black points). Similarly, the summed resolved SEDs that include outskirt components are nearly identical to the unresolved SEDs obtained from the 3D-HST photometry (red and black curves).\footnote{As stated at the beginning of \autoref{sec:integrated}, small differences between the total flux and the 3D-HST photometry are expected, but generally agree within the uncertainties.} Note that not all galaxies, such as GOODS-N\,21507, have an outskirt component as all of the regions have sufficient S/N to be included as Voronoi bins. 

From \autoref{fig:sedcomp}, we conclude that the \textit{shapes} of the SEDs are generally well-preserved between the resolved and integrated results (with the exception of galaxies similar to AEGIS\,8907, which are further discussed in \autoref{sec:intparam}). The primary difference between the resolved and integrated SEDs is the \textit{normalization}, which is attributed to the fact that the Voronoi bins contain less light than the broadband measurements. This causes the Voronoi bins to exhibit systematically lower stellar masses and SFRs compared to the 3D-HST photometry (as expected), but on average the offset is less than the uncertainties in the SED-derived stellar masses and SFRs. 
\begin{figure*}
\begin{adjustbox}{width=\linewidth, center}
\begin{minipage}{\linewidth}
\includegraphics[width=\linewidth]{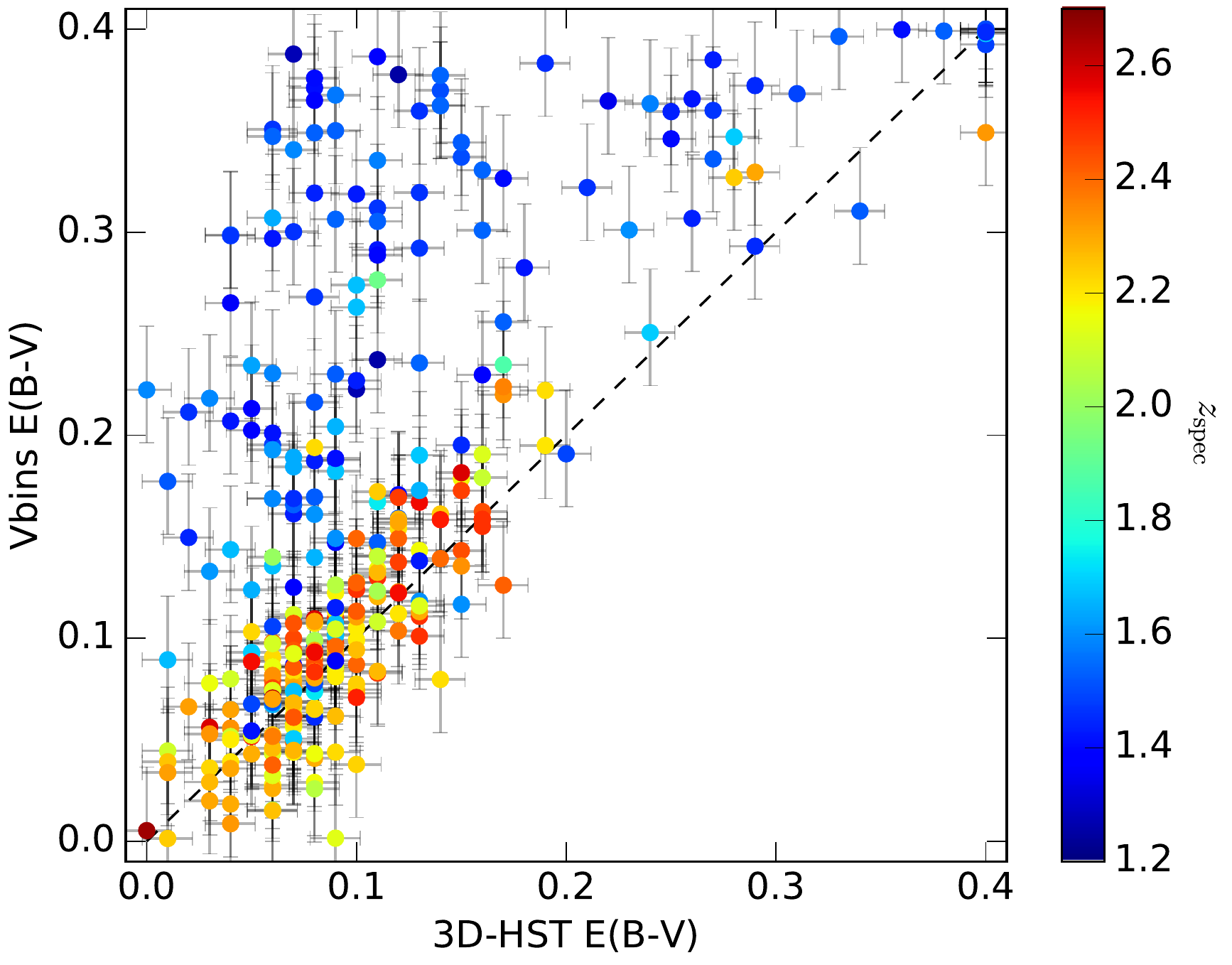}
\end{minipage}
\quad
\begin{minipage}{\linewidth}
\includegraphics[width=\linewidth]{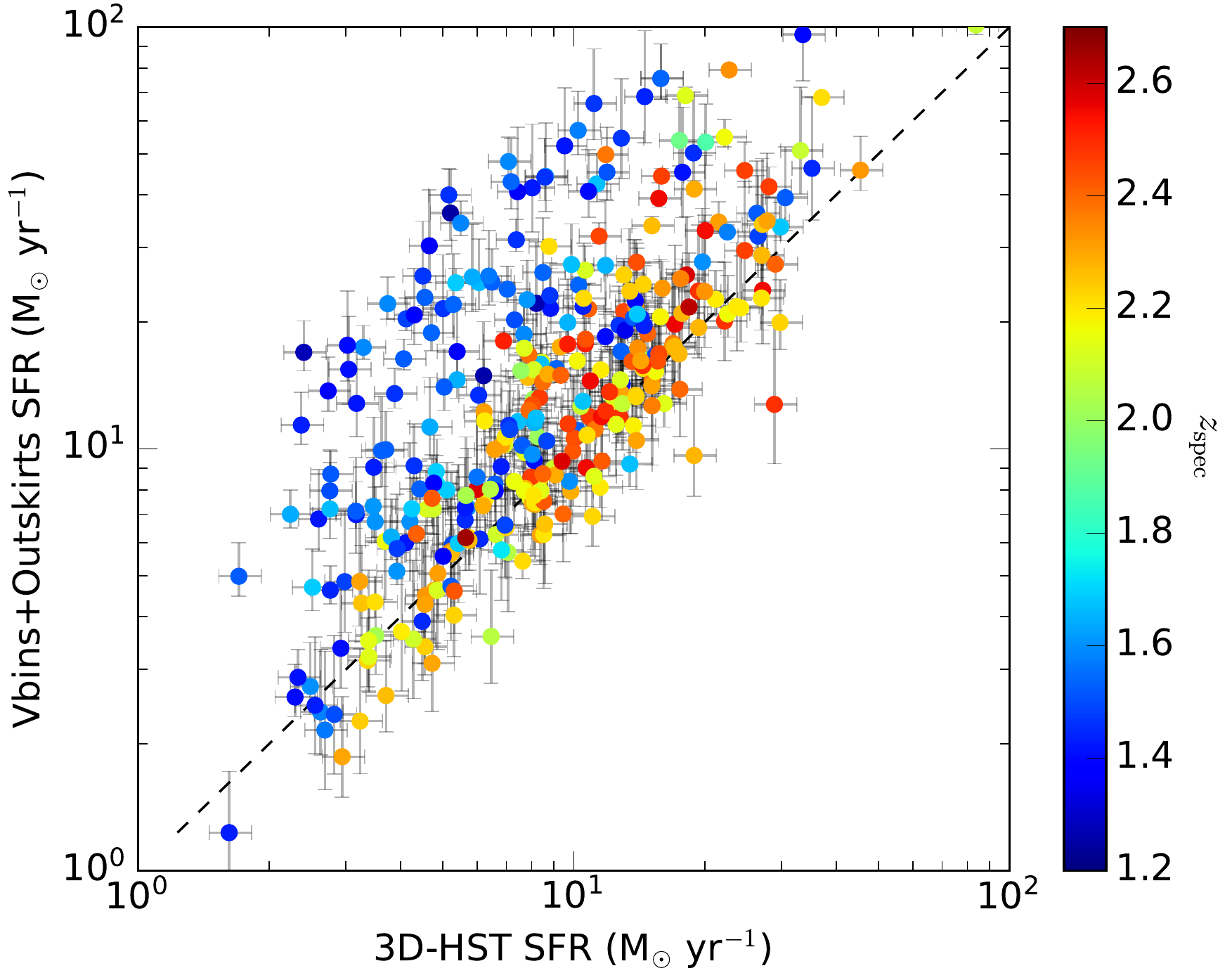}
\end{minipage}
\end{adjustbox}
\caption{\textit{Left:} The average Voronoi bin \EBVs\ weighted by the area of the Voronoi bins versus the \EBVs\ derived from the 3D-HST photometry. \textit{Right:} The total SFR of the Voronoi plus outskirt bins versus the SFR derived from the 3D-HST photometry. 
The points are color-coded by their spectroscopic redshift. The dashed black line indicates where the \EBVs\ or SFR derived from the Voronoi bins equals those inferred from the 3D-HST photometry. Only the galaxies in the $z\sim1.5$ sample exhibit \EBVs\ or SFRs that deviate significantly from those derived from the 3D-HST photometry due to the lack of resolved imaging covering the UV slope (e.g., AEGIS\,8907 in \autoref{fig:sedcomp}).}
\label{fig:pcomp_z}
\end{figure*}

\subsection{Comparison of SED-derived Parameters}\label{sec:intparam}
\begin{figure*}
\begin{adjustbox}{width=\linewidth, center}
\begin{minipage}{\linewidth}
\includegraphics[width=\linewidth]{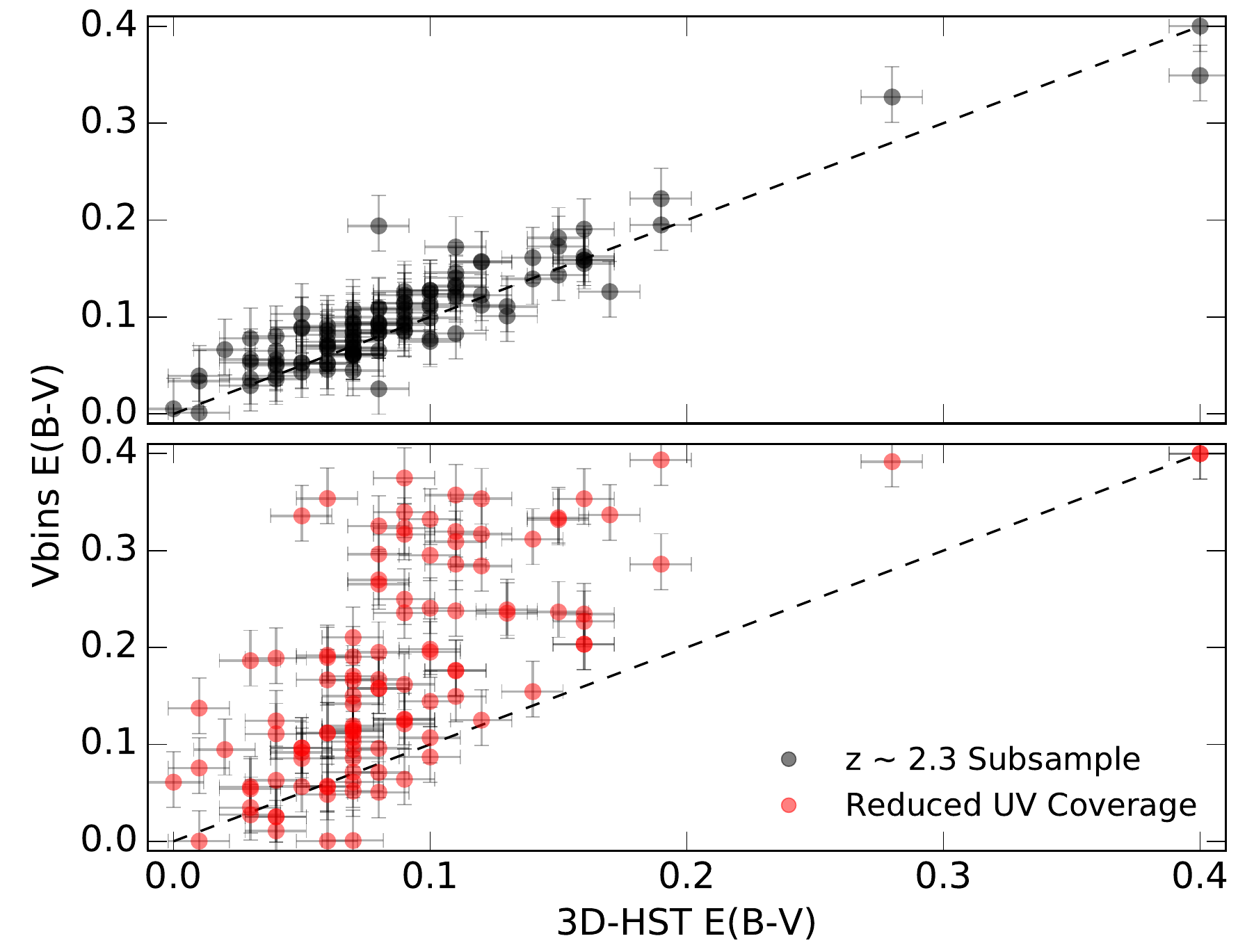}
\end{minipage}
\quad
\begin{minipage}{\linewidth}
\includegraphics[width=\linewidth]{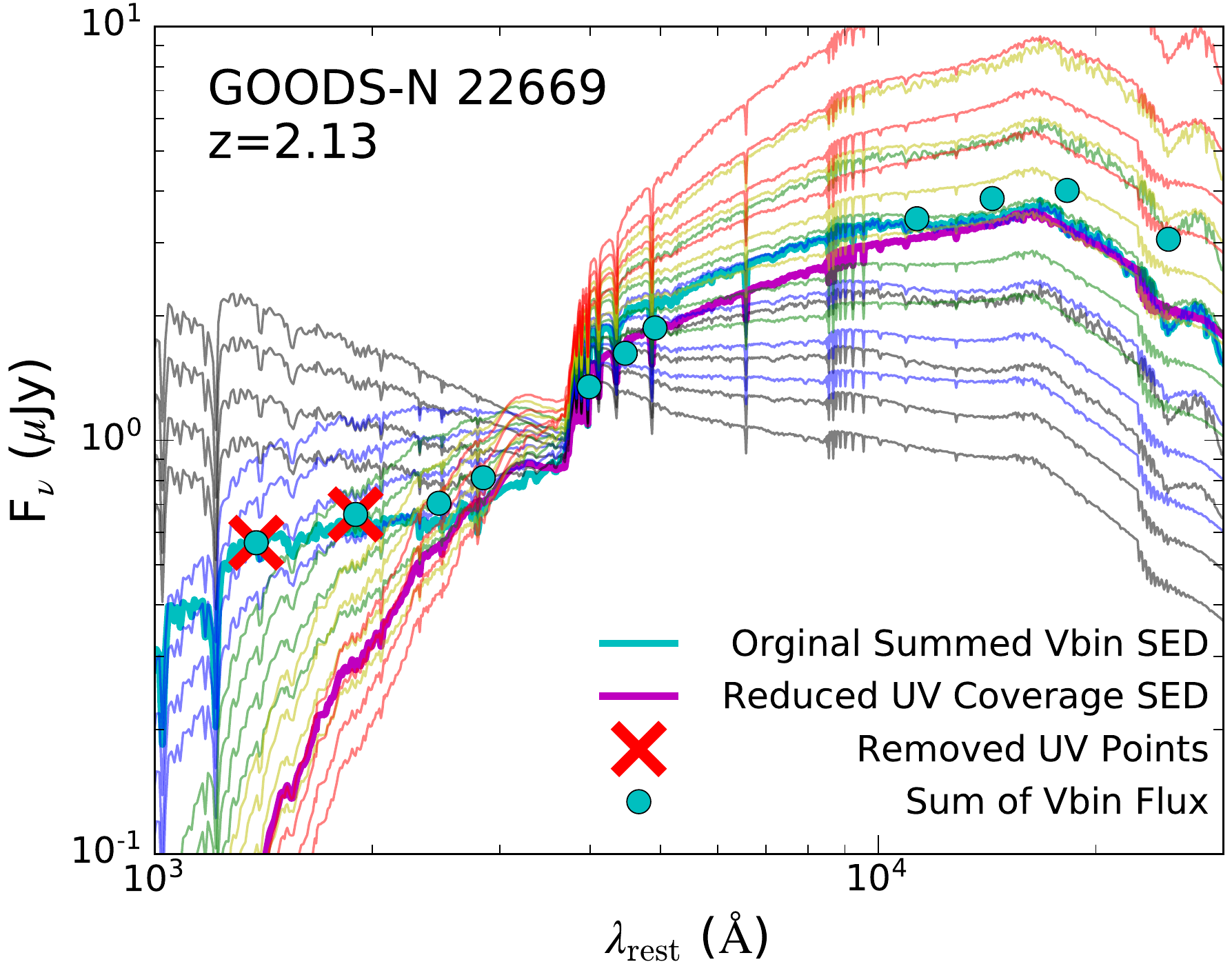}
\end{minipage}
\end{adjustbox}
\caption{\textit{Left:} The average Voronoi bin \EBVs\ weighted by the area of the Voronoi bins versus the \EBVs\ derived from the 3D-HST photometry for the $z\sim2.3$ galaxies. The black points show the original \EBVs\ of the $z\sim2.3$ galaxies (\textit{top}). The red points show how removing the two bluest filters from the $z\sim2.3$ galaxies reproduces the relative reddening observed in the $z\sim1.5$ sample (\textit{bottom}). \textit{Right:} The sum of the model SEDs that best fit the resolved Voronoi bins for GOODS-N\,22669 ($z=2.13$). The thick light blue curve shows the summed SED that includes all of the data from the resolved imaging (blue points). The thick purple curve shows the summed SED where the data from the two bluest filters were not included in the SED fitting (red crosses). The faded curves in black, blue, green, yellow, and red represent SED models with \EBVs\ equal to 0.0, 0.1, 0.2, 0.3, and 0.4, respectively. Several curves of the same color are shown to demonstrate how the SED model shape varies with different stellar population ages when \EBVs\ is held constant.}
\label{fig:sedmod_compare}
\end{figure*}
The shapes of the resolved SEDs are generally consistent with the unresolved SEDs presented (\autoref{fig:sedcomp}), but several galaxies exhibit summed resolved SEDs that are systematically offset from the integrated SED. The \EBVs\ and stellar population age derived from the SED that is best-fit to the integrated 3D-HST photometry is compared to the average \EBVs\ and stellar population age derived from the Voronoi bins weighted by the area of each Voronoi bin.\footnote{Noisy outskirt bins that do not satisfy a S/N~$\geq5$ in at least 5 filters are generally not included in the analysis unless specified (see \autoref{sec:vbin}). Therefore, the area-weighted average \EBVs\ and stellar population age of the Voronoi bins for an individual galaxy is not biased towards large outskirt bins with poorly constrained photometry. Furthermore, alternatively weighting individual Voronoi bins by either the $H_{160}$ flux or log stellar mass results in average \EBVs\ and stellar population ages that are within 1$\sigma$ of the area-weighted average \EBVs\ and stellar population ages, such that the results presented here do not change with this alternate treatment.} The stellar masses and SFRs derived from the SED fit to the integrated photometry is compared to the sum of the stellar mass and SFR within all of the Voronoi bins. While the SED normalization differences caused by different area coverage between the Voronoi bins and integrated photometry do cause systematically lower stellar masses and SFRs, we find that on average all SED-derived parameters inferred from the resolved SEDs are within one standard deviation of those derived from the integrated 3D-HST photometry. 

A notable exception is that the resolved SEDs of the $z\sim1.5$ galaxies are on average redder and younger than indicated by the broadband SED fits, as is showcased by AEGIS\,8907 in \autoref{fig:sedcomp}. The AEGIS\,8907 photometry demonstrates that this discrepancy is caused by the lack of resolved imaging covering the rest-frame UV for our $z\sim1.5$ sample. The number of filters that cover the UV slope depends on the field, where all of the $z\sim1.5$ galaxies in AEGIS, COSMOS, or UDS have only one {\em HST} filter available that covers the UV slope (1250--2500\,\AA). We find that the best-fit SEDs prefer redder and younger solutions when the UV slope is unconstrained. Only above redshifts $z\gtrsim2.1$ do AEGIS, COSMOS, and UDS galaxies have at least two filters covering the UV slope. The left panel of \autoref{fig:pcomp_z} demonstrates how the resolved SED fitting of the $z\sim1.5$ sample produces redder inferred \EBVs\ compared to those from the unresolved broadband SED fits, while the preference for redder \EBVs\ completely disappears for the $z\sim2.3$ sample. Similarly, the right panel of \autoref{fig:pcomp_z} shows how the redder \EBVs\ and younger stellar ages in the $z\sim1.5$ sample causes the derived SFRs from the Voronoi plus outskirt bins\footnote{SFRs inferred from the Voronoi bins alone are lower than those inferred from the 3D-HST photometry because they cover a smaller region of the imaging. Therefore, comparing the SFRs derived from the Voronoi plus outskirt bins is more appropriate for the right panel of \autoref{fig:pcomp_z}.} to be overestimated compared to SFRs inferred from the 3D-HST photometry. The discrepancy between the inferred \EBVs\ derived from the unresolved and resolved photometry in the $z\sim1.5$ sample occurs at all \EBVs\ values that are probed by the SED model grid ($0.0\leq \EBVs \leq0.4$) and does not correlate with stellar mass, such that the measured differences in \EBVs\ are not likely associated with our choice in dust attenuation curve (see \autoref{fn:atten_curve}). However, there may be intrinsic differences in the stellar populations between the two redshift samples.

In order to further investigate whether the redder inferred \EBVs\ are caused by the lack of UV slope coverage or differences between the intrinsic stellar populations in the two redshift bins, we select 111~galaxies from our $z\sim2.3$ sample that have observations in at least 7 resolved \HST\ filters. The two bluest filters observed are removed such that only one remaining filter covers the UV slope, with at least 5 resolved filters remaining for the SED fitting. As can be seen in the right panel of \autoref{fig:pcomp_z}, when the resolved SED fitting (see \autoref{sec:sedfit_res}) is repeated we find that the average \EBVs\ inferred across all Voronoi bins is redder than that inferred when there are additional filters covering the UV slope. Therefore, we conclude that the inferred \EBVs\ values for the $z\sim1.5$ sample are overestimated due to there being fewer than 2 filters covering the UV slope. \autoref{fig:sedmod_compare} shows how the sum of the resolved SEDs changes (thick light blue curve compared to the thick purple curve) when the 2 bluest filters covering the UV slope are removed (red crosses) for GOODS-N\,22669 ($z=2.13$). For context, a subset of the SED models are also shown, where each color represents a step in \EBVs\ for a range of stellar population ages (i.e., the black, blue, green, yellow, and red SEDs represent SED models with $0\leq\EBVs\leq0.4$ in steps of 0.1\,mag, respectively). There are three significant features that can be seen in \autoref{fig:sedmod_compare}: 1) when the UV photometry is included, the shape of the best-fit SED is primarily constrained by the rest-frame UV photometry (light blue curve); 2) when the UV photometry is not included, the shape of the best-fit SED is primarily constrained by the rest-frame optical photometry (purple curve); and 3) the SED fits generally under-predict the IRAC photometry. The IRAC photometry may be brighter than what can be fit with the model SEDs in the rest-frame near-IR due to the higher sensitivity and larger aperture over which these measurements are made.  Systematically brighter IRAC fluxes and the known discrepancy between stellar population age and extinction in the SED models at optical wavelengths \citep[e.g.,][]{Worthey94, Shapley01} may together explain why there is a bias towards redder best-fit SEDs when the UV slope is unconstrained (such as AEGIS\,8907). 

The lack of UV slope coverage in the resolved Voronoi bin fitting for the $z\sim1.5$ sample may complicate inferences of the intrinsic variation in the reddening distribution for these galaxies. However, rather than observing a flat reddening distribution in the $z\sim1.5$ galaxies, structure is observed in the \EBVs\ maps (e.g., see \autoref{fig:map_ex} for UDS\,21834 at $z=1.67$) due to the reddening and ages being constrained in part by the strength of the Balmer/4000\,\AA~breaks in the SED fitting. Finally, we note that the SED-derived SFRs and stellar masses are constrained by the overall normalization of the SED, and they are minimally affected by the lack of observations covering the UV slope. 

\subsection{Traditional Voronoi Binning}\label{sec:traditional}
\begin{figure*}
\begin{adjustbox}{width=\linewidth, center}
\begin{minipage}{\linewidth}
\includegraphics[width=\linewidth]{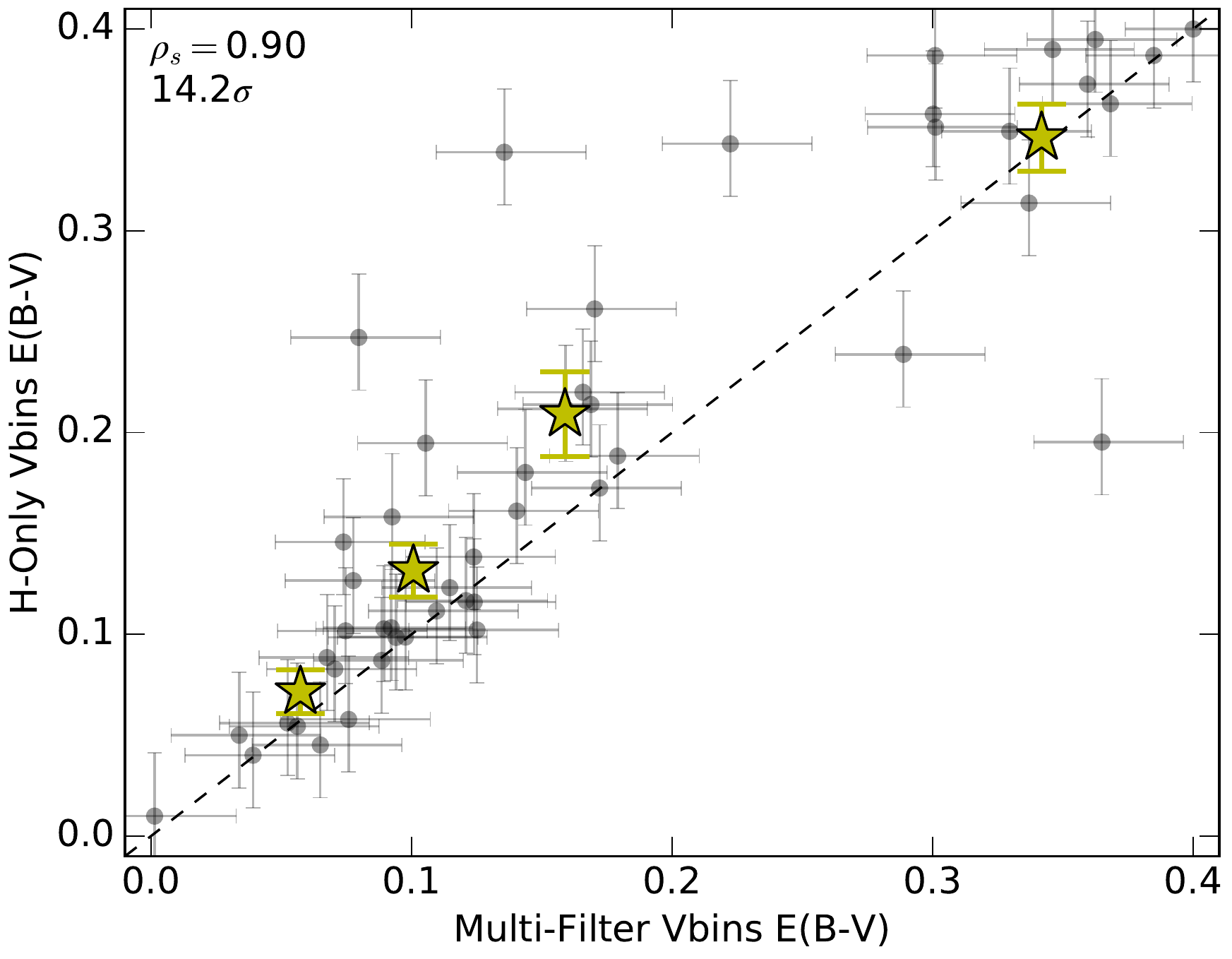}
\end{minipage}
\quad
\begin{minipage}{\linewidth}
\includegraphics[width=\linewidth]{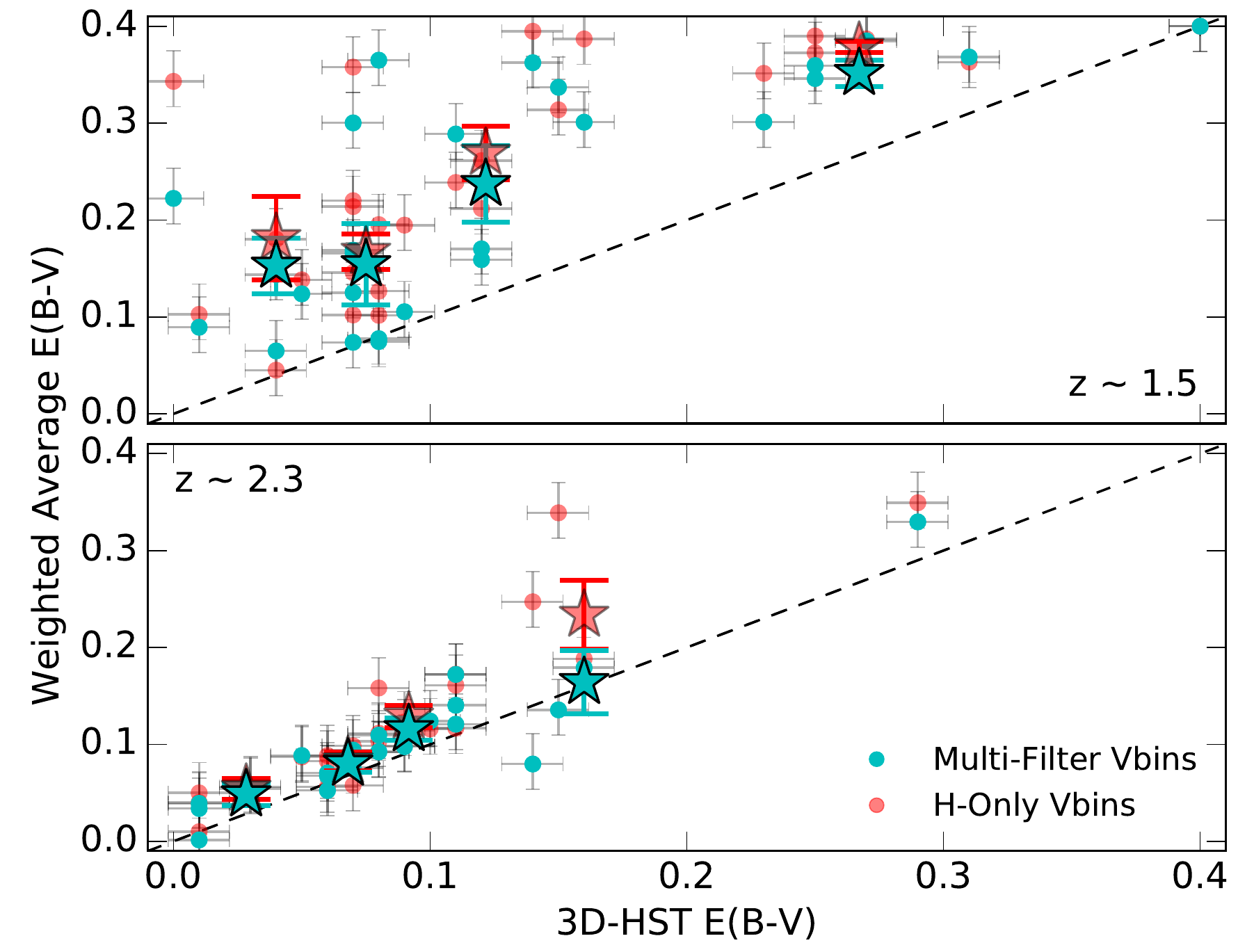}
\end{minipage}
\end{adjustbox}
\caption{\textit{Left:} The average \EBVs\ weighted by the area of the Voronoi bins derived from the ``traditional'' Voronoi bins derived from the $H_{160}$ S/N distribution exclusively versus the Voronoi bin distribution derived from several filters (see \autoref{sec:vbin}). The Spearman rank correlation coefficient and its significance are listed in the top left corner. \textit{Right:} The average \EBVs\ weighted by the area of the Voronoi bins derived from both Voronoi binning schemes compared to the \EBVs\ inferred from the 3D-HST photometry for both redshift samples. The red points represent the ``traditional'' Voronoi binning technique and the light blue points represent our multi-filter Voronoi binning technique. The red and light blue stars are the averages from individual points equally divided into four bins of \EBVs\ in the x-axis. The black dashed lines indicate where the \EBVs\ derived from all sources are equal.}
\label{fig:pcomp_Honly}
\end{figure*}
In constructing the Voronoi bins in \autoref{sec:vbin}, we enforce more stringent requirements (minimum 3 pixels and S/N~$\ge5$ in 5 filters) than the ``traditional'' Voronoi binning procedure (S/N~$\ge10$ in $H_{160}$) performed in previous studies \citep{Wuyts12, Wuyts13, Wuyts14, Genzel13, Tadaki14, Lang14}. In this section, the results from our multi-filter Voronoi binning technique is directly compared to the traditional Voronoi binning approach using the same subset of 50~galaxies that were used to determine the typical SED errors (see \autoref{sec:sedfit} and \autoref{sec:sedfit_res}). Only 50~galaxies are selected for this analysis in the interest of limiting computation time, but the range of SED parameters probed by these 50 galaxies is similar to that of the larger sample. The traditional Voronoi bins are constructed using the original \citet{Cappellari03} algorithm (version 3.0.4), which exclusively requires each Voronoi bin to attain a S/N~$\geq10$ in the $H_{160}$ filter alone. Traditional Voronoi bin construction failed for one galaxy, resulting in a final subsample of 49 galaxies for this exercise.

The left panel of \autoref{fig:pcomp_Honly} shows that the average \EBVs\ inferred across the traditional Voronoi bins determined by the $H_{160}$-band alone is typically redder (higher) compared to the average \EBVs\ inferred from the multi-filter determined Voronoi bins. For context, the average \EBVs\ derived from both Voronoi binning procedures are also compared to the \EBVs\ derived from the unresolved 3D-HST photometry in the right panel of \autoref{fig:pcomp_Honly}. We find that the average \EBVs\ inferred from the traditional $H_{160}$-only Voronoi bins (red points and stars) are on average 0.02\,mag redder (higher) than our multi-filter approach (light blue points and stars) in both redshift bins, but deviate as much as 0.20\,mag. Meanwhile, the \EBVs\ from our multi-filter approach are 0.10 and 0.02\,mag redder (higher) than the 3D-HST globally derived \EBVs\ (black dashed line) for our $z\sim1.5$ and $z\sim2.3$ samples, respectively. For the $z\sim1.5$ sample, the likely explanation as to why both resolved methods exhibit redder (higher) \EBVs\ compared those derived from the unresolved 3D-HST photometry is that the UV slope is unconstrained by the available resolved data, as explained in \autoref{sec:intparam}. Meanwhile, the differences between the \EBVs\ derived from our multi-filter technique and the unresolved 3D-HST photometry are on average smaller than the differences in the \EBVs\ derived from the traditional Voronoi binning technique and the unresolved 3D-HST photometry. This is due to the larger Voronoi bin sizes inherent to our technique that are consequently less resolved compared to Voronoi bins that are defined by a single filter. 

\begin{figure*}
\includegraphics[width=.9\textwidth]{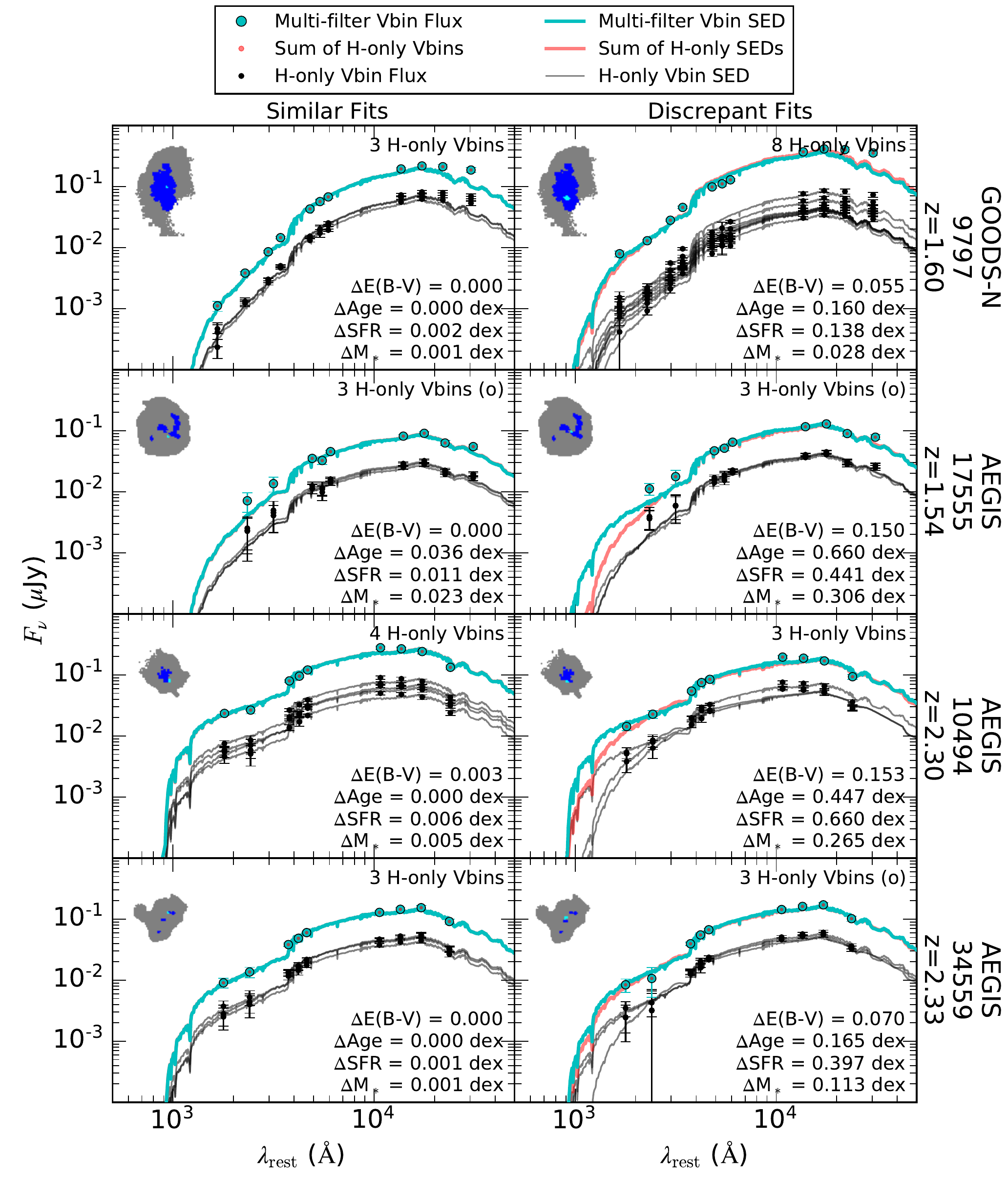}
\caption{Examples of the best-fit SEDs from the Voronoi bins built from the $H_{160}$ filter alone compared to best-fit SED of the single Voronoi bin constructed from our multi-filter procedure (\autoref{sec:vbin}) that covers the same physical area as the $H_{160}$-only Voronoi bins. The black points represent the fluxes within the $H_{160}$-only Voronoi bins, which add up to the fluxes from a single, larger Voronoi bin built using our new criteria (light blue/red points). The errors for the light blue/red points are typically smaller than the size of the points. The gray curves are the SEDs that are best fit to the $H_{160}$-only Voronoi bin fluxes. The red curves represent the sum of the gray curves and are generally hidden behind the light blue curves that represent the SEDs that are best fit to the individual Voronoi bin constructed from our multi-filter procedure. The number of $H_{160}$-only Voronoi bins is listed in the top right corner of each panel and an (o) is shown if the matched multi-filter Voronoi bin is considered an outskirt component. The values in the bottom right corner of each panel show the differences in \EBVs, stellar population age, SFR, and stellar mass inferred from the integrated $H_{160}$-only Voronoi bin SEDs versus the individual Voronoi bin from our multi-filter procedure. The image in the top left corner of each panel shows for each galaxy: the segmentation map area in gray, the area where the Voronoi bins have a S/N~$\geq5$ in at least 5 filters in dark blue, and the individual Voronoi bin that is represented by the curves in light blue. The two columns show examples where the summed SEDs (light blue and red curves) are similar (\textit{left column}) or discrepant (\textit{right column}) compared to each other.}
\label{fig:vorHonly_vbin}
\end{figure*}

The differences observed between the \EBVs\ inferred from the $H_{160}$ derived Voronoi bins and the multi-filter derived Voronoi bins could also be explained by the high S/N requirement for bluer filters in the multi-filter technique, in that it will force blue regions of the galaxy to be included with intrinsically red regions. Forcing additional blue light to be included with intrinsically red regions would cause the multi-filter Voronoi bins to typically exhibit systematically shallower UV slopes compared to traditionally defined Voronoi bins---particularly those covering the reddest regions of the galaxy. Therefore, in \autoref{fig:vorHonly_vbin}, the resolved SEDs that cover the same spatial areas from the two methodologies are directly compared such that they have the same total fluxes (light blue and red points). Individual Voronoi bins built from the multi-filter procedure (\autoref{sec:vbin}) are typically larger than traditionally constructed Voronoi bins based only on the $H_{160}$ filter. Specifically, \autoref{fig:vorHonly_vbin} shows examples of the resolved SEDs from the traditional Voronoi bins (black points and gray curves) that entirely make up an individual Voronoi bin from the multi-filter procedure (light blue points and curves). As can be seen, there are examples where the shape of the summed SEDs derived from the traditionally defined $H_{160}$-only Voronoi bins (light blue curves) either matches (left column) or deviates (right column) from the shape of the SEDs derived from the multi-filter Voronoi bins covering the same region (red curves). Based on the matching UV slopes from the examples shown in the left column of \autoref{fig:vorHonly_vbin} and the fact that the discrepant SED examples in the right column have been compared over the same resolved regions where all filters contain the same total flux, we conclude that bluer light being forced into the multi-filter Voronoi bins does not serve as an adequate explanation to the observed systematic difference between the \EBVs\ inferred from the two methodologies in \autoref{fig:pcomp_Honly}. However, \autoref{fig:vorHonly_vbin} also shows that the best-fit SEDs from the multi-filter technique are typically more tightly constrained by the higher S/N photometry than the SEDs fit to the smaller Voronoi bins constructed from the $H_{160}$ filter alone. The inclusion of any resolved components with low S/N at short wavelengths appears to cause the summed best-fit SEDs to exhibit steeper UV slopes that result in redder (higher) \EBVs\ for galaxies in both the $z\sim1.5$ and $z\sim2.3$ samples (right column of \autoref{fig:vorHonly_vbin}). Furthermore, since the discrepancy between the UV slopes of the two Voronoi binning methods is observed in both of our redshift bins, it is important to note that the redder inferred \EBVs\ from the summed traditional SEDs is independent from the redder \EBVs\ that are caused by the lack of resolved UV slope coverage in the $z\sim1.5$ sample (see \autoref{sec:intparam}). Overall, the preference towards redder SEDs tends to occur when the UV slope is generally poorly constrained (in this case by large photometric errors or, as discussed in \autoref{sec:intparam}, limited data). 

Based on the results presented here, we find that traditionally defined Voronoi bins result in resolved SEDs that are \textit{systematically redder} (higher \EBVs) than multi-filter defined Voronoi bins. This systematic effect is due to the lower S/N in bluer filters when constructing Voronoi bins based only on the distribution of $H_{160}$-band light. By extension, we expect SED fitting across individual pixels to exhibit redder (higher) \EBVs\ as well. In order to best constrain the shape of the SEDs that are fit to resolved photometry, we recommend using a binning procedure that enforces high S/N across both blue and red resolved filters---especially covering the UV slope and Balmer/4000\,\AA\ breaks. Additionally, caution should be taken when including regions that have low S/N overall (i.e., ``outskirt'' components) in resolved studies. 

\subsection{Discussion}\label{sec:discussion}
The structure of high-redshift galaxies can give insight into the evolution of galaxies on resolved scales \citep[see][]{Conselice14}. In particular, the SFR of a galaxy is an important measure of how galaxies build their stellar mass. However, the intrinsic SFR of a galaxy must be inferred by correcting for the obscuring effects of dust. The scattering and attenuation of starlight by dust depends on the amount and distribution of dust \citep[e.g.,][]{Draine84, Fitzpatrick86, Calzetti94, Calzetti00}, which is typically accounted for by assuming an attenuation curve. Obtaining accurate dust reddening on resolved scales may inform the assumptions made when fitting unresolved photometry, such as the slope and shape of the dust attenuation curve \citep{Reddy18-1, Shivaei20}. While stellar masses are generally quite robust to methodology, we have shown throughout this paper that the inferred resolved reddening can be difficult to constrain due to a combination of the age-extinction-metallicity degeneracy (\citealt{Worthey94, Shapley01}; see \aref{app:colors}), unconstrained UV slopes (see \autoref{sec:intparam}), and generally low S/N resolved rest-frame UV photometry (see \autoref{sec:traditional}). In such cases, Voronoi bins constructed based on multiple filters may be desirable in order to obtain more robust resolved stellar population parameters and, in particular, robust reddening maps.

In \aref{app:colors}, we discussed how constraining the Balmer/4000\,\AA\ breaks is important for breaking the age-extinction degeneracy \citep{Worthey94, Shapley01}, particularly for galaxies at $z>2.5$ where the Balmer/4000\,\AA\ breaks shift into the $H_{160}$ filter. Through our choice of normalizing the unresolved IRAC photometry by the resolved $H_{160}$ flux, we made the assumption that the resolved \HIRAC\ colors are constant across all bins in each galaxy. We showed that constraining the \HIRAC\ colors does not significantly influence the distribution of stellar population ages and reddening within galaxies. While only 15~galaxies in our sample are at $z>2.5$, constraining the Balmer/4000\,\AA\ breaks by assuming a constant \HIRAC\ color distribution may be a viable solution for resolved studies at higher redshifts. 

Throughout this paper, we showed that if the UV slope is not well-constrained by the resolved photometry, such as at lower redshifts ($z<2.1$; see \autoref{sec:intparam}) or by low S/N regions (see \autoref{sec:traditional}), then the derived resolved stellar populations properties may be biased towards SEDs with redder (higher) \EBVs. If the inferred \EBVs\ is overestimated, then observed SFRs may consequently be overcorrected for dust attenuation, which would result in the inferred intrinsic SFRs also being overestimated (see \autoref{fig:pcomp_z}). This observational reddening bias most significantly affects our $z\sim1.5$ sample, where only one resolved filter covers the UV slope (see \autoref{sec:intparam}), but more generally this systematic bias has a greater influence on intrinsically blue galaxies (young, dust-free stellar populations) opposed to intrinsically red galaxies (old, dusty stellar populations). A \citet{Wuyts12} methodology can be implemented to adjust the resolved SEDs to better reproduce the integrated SED shape in cases where it is not possible to obtain at least two resolved filters covering the UV slope (1250--2500\,\AA), but at the cost of increased uncertainty in the specific distribution of the resolved stellar population properties (see \aref{app:wuyts}). For galaxies where the resolved UV slope is not well-constrained due to low S/N, we introduced a technique in \autoref{sec:vbin} that increases the S/N across all resolved filters to produce more reliable SEDs that are fit to the resolved photometry at the cost of a marginally decreased resolution compared to previous methods (see \autoref{sec:traditional}). Our multi-filter Voronoi binning technique increases the number of resolved elements that can confidently recover resolved stellar population parameters through SED fitting, particularly for galaxies with low surface brightness or extended diffuse light. 

Based on the analyses presented throughout this paper, we provide a list of recommendations for confidently deriving resolved stellar population properties. 1) If possible, ensure that at least two photometric observations span key rest-frame SED features, particularly the UV slope and Balmer/4000\,\AA\ breaks. 2) Incorporate a binning technique that either includes information about the S/N distribution of both red and blue filters or otherwise ensures that at least 5~filters are significantly detected---especially shorter wavelength filters that are typically lower S/N. 3) Generally be wary of biases that may exist between resolved properties that are inferred from well- (e.g., full wavelength coverage, S/N $\geq5$ in 5 filters) versus poorly-constrained (e.g., unconstrained SED features, ``outskirt'' components) SEDs. 

\section{Summary}\label{sec:summary}
In this paper, we presented an improved Voronoi binning technique for studying resolved stellar populations at high redshifts, which overall aims to increase the confidence in the best-fit SEDs and stellar population properties that are derived from the resolved photometry. The layout of the resolved components for performing resolved SED fitting was determined from the S/N distribution of multiple filters in the high-resolution CANDELS/3D-HST imaging for a sample of \nsamp~star-forming galaxies drawn from the MOSDEF survey at spectroscopic redshifts $\zmin<z<\zmax$. The stellar population and dust maps were constructed based on our new multi-filter Voronoi binning technique and their integrated stellar population properties were compared to those derived from the unresolved broadband imaging and those derived from a ``traditional'' Voronoi binning approach that used the $H_{160}$ filter alone to determine Voronoi bins. Our main conclusions are summarized as follows. 

\begin{itemize}

\item In \autoref{sec:vbin}, we described a modified Voronoi binning technique (see \autoref{fig:vbin_flow}) that accounts for the S/N distribution of multiple resolved filters opposed to the $H_{160}$ filter alone. In order for an individual Voronoi bin to be included in our analysis, we required the bin to contain a S/N $\geq5$ in at least 5 resolved filters. Any bins that did not satisfy this criteria were deemed ``outskirt'' components (see \autoref{fig:vbin_steps}).

\item We found that the shapes of the summed resolved SEDs from our multi-filter binning technique were generally consistent with the SEDs that were fit to the unresolved broadband photometry (see \autoref{fig:sedcomp}) in \autoref{sec:intsed}. Similarly, in \autoref{sec:intparam} we found that the average \EBVs, average stellar population ages, summed SFRs, and summed stellar masses are typically within $1\sigma$ of the stellar population properties derived from the unresolved broadband photometry. 

\item The most significant deviation between the resolved and unresolved SEDs and best-fit stellar population parameters was the shape of the UV slope and the average \EBVs\ inferred from the resolved reddening maps for galaxies in the $z\sim1.5$ sample (see \autoref{fig:pcomp_z}). By removing rest-frame UV photometry from the $z\sim2.3$ sample and repeating the resolved SED fitting, we determined in \autoref{sec:intparam} that the discrepancy in the $z\sim1.5$ inferred reddening is most likely caused by the lack of data that is able to constrain the UV slope (1250--2500\,\AA). 

\item When comparing our multi-filter Voronoi binning technique (S/N $\geq5$ in at least 5 filters) with a ``traditional'' single-filter approach (S/N $\geq10$ in $H_{160}$) in \autoref{sec:traditional}, we found that the single-filter technique results in \EBVs\ values that are \textit{systematically redder} by 0.02\,mag on average than our multi-filter method, but could deviate by as much as 0.20\,mag (see \autoref{fig:pcomp_Honly}). We also found that our multi-filter technique produces average inferred \EBVs\ that are closer to those derived from the unresolved broadband photometry compared to the single-filter approach, but we note that this may be expected from the larger (i.e., lower resolution) bins that are inherent to our technique. We also demonstrated through \autoref{fig:vorHonly_vbin} that systematically redder UV slopes may result from poorly constrained rest-frame UV photometry, which can occur in the smaller, traditionally defined Voronoi bins. Finally, we note that SED-derived stellar masses remain robust between methodologies when all Voronoi bins are included. 

\end{itemize}

The methodology presented here will be applied to future resolved analyses using the MOSDEF dataset, which will additionally incorporate results from the rest-frame optical emission line measurements. In general, we advocate for a methodology that best constrains the SEDs that are best-fit to the resolved photometry. Well-constrained SEDs can generally be obtained by using high S/N multi-band photometry that covers key features in the rest-frame SED (i.e., UV slope, Balmer/4000\,\AA\ breaks). Furthermore, low S/N ``outskirt'' regions could be excluded from analyses, otherwise biases between well- versus poorly-constrained resolved SEDs must be understood when interpreting the results. Finally, we note that higher confidence in resolved stellar population parameters through the methods presented here comes at the cost of lower resolution. However, these methods will similarly push to higher resolutions when applied to future high-resolution imaging instruments, such as the NIRcam on the upcoming \textit{James Webb Space Telescope}. 

\section*{Acknowledgements}
This work is based on observations taken by the CANDELS Multi-Cycle Treasury Program and the 3D-HST Treasury Program (GO 12177 and 12328) with the NASA/ESA \HST, which is operated by the Association of Universities for Research in Astronomy, Inc., under NASA contract NAS5-26555. The MOSDEF team acknowledges support from an NSF AAG collaborative grant (AST-1312780, 1312547, 1312764, and 1313171) and grant AR-13907 from the Space Telescope Science Institute. The authors wish to recognize and acknowledge the very significant cultural role and reverence that the summit of Maunakea has always had within the indigenous Hawaiian community. We are most fortunate to have the opportunity to conduct observations from this mountain.
\\ \\
\textit{Facilities:} HST (WFC3, ACS), Keck:I (MOSFIRE), Spitzer (IRAC)
\\ \\
\textit{Software:} Astropy \citep{Astropy_collaboration13, Astropy_collaboration18}, Matplotlib \citep{Hunter07}, NumPy \citep{Oliphant07}, SciPy \citep{Oliphant07}, Voronoi Binning Method \citep{Cappellari03}

\section*{Data Availability}
Resolved CANDELS/3D-HST photometry is available at: \url{https://3dhst.research.yale.edu/Data.php}. Spectroscopic redshifts from the MOSDEF survey are available at: \url{http://mosdef.astro.berkeley.edu/for-scientists/data-releases/}.
%
%
%

%\bibliographystyle{mnras}
%\bibliography{mapsbib}

\begin{thebibliography}{}
\makeatletter
\relax
\def\mn@urlcharsother{\let\do\@makeother \do\$\do\&\do\#\do\^\do\_\do\%\do\~}
\def\mn@doi{\begingroup\mn@urlcharsother \@ifnextchar [ {\mn@doi@}
  {\mn@doi@[]}}
\def\mn@doi@[#1]#2{\def\@tempa{#1}\ifx\@tempa\@empty \href
  {http://dx.doi.org/#2} {doi:#2}\else \href {http://dx.doi.org/#2} {#1}\fi
  \endgroup}
\def\mn@eprint#1#2{\mn@eprint@#1:#2::\@nil}
\def\mn@eprint@arXiv#1{\href {http://arxiv.org/abs/#1} {{\tt arXiv:#1}}}
\def\mn@eprint@dblp#1{\href {http://dblp.uni-trier.de/rec/bibtex/#1.xml}
  {dblp:#1}}
\def\mn@eprint@#1:#2:#3:#4\@nil{\def\@tempa {#1}\def\@tempb {#2}\def\@tempc
  {#3}\ifx \@tempc \@empty \let \@tempc \@tempb \let \@tempb \@tempa \fi \ifx
  \@tempb \@empty \def\@tempb {arXiv}\fi \@ifundefined
  {mn@eprint@\@tempb}{\@tempb:\@tempc}{\expandafter \expandafter \csname
  mn@eprint@\@tempb\endcsname \expandafter{\@tempc}}}

\bibitem[\protect\citeauthoryear{Abraham, van~den Bergh  \& Nair}{Abraham
  et~al.}{2003}]{Abraham03}
Abraham R.~G.,  van~den Bergh S.,   Nair P.,  2003, \mn@doi [ApJ]
  {10.1086/373919}, 588, 218

\bibitem[\protect\citeauthoryear{{Astropy Collaboration} et~al.,}{{Astropy
  Collaboration} et~al.}{2013}]{Astropy_collaboration13}
{Astropy Collaboration} et~al., 2013, \mn@doi [A\&A]
  {10.1051/0004-6361/201322068}, 558, A33

\bibitem[\protect\citeauthoryear{{Astropy Collaboration} et~al.,}{{Astropy
  Collaboration} et~al.}{2018}]{Astropy_collaboration18}
{Astropy Collaboration} et~al., 2018, \mn@doi [AJ] {10.3847/1538-3881/aabc4f},
  156, 123

\bibitem[\protect\citeauthoryear{Azadi et~al.,}{Azadi et~al.}{2017}]{Azadi17}
Azadi M.,  et~al., 2017, \mn@doi [ApJ] {10.3847/1538-4357/835/1/27}, 835, 27

\bibitem[\protect\citeauthoryear{Azadi et~al.,}{Azadi et~al.}{2018}]{Azadi18}
Azadi M.,  et~al., 2018, \mn@doi [ApJ] {10.3847/1538-4357/aad3c8}, 866, 63

\bibitem[\protect\citeauthoryear{Barden, Jahnke  \& H{\"a}u{\ss}ler}{Barden
  et~al.}{2008}]{Barden08}
Barden M.,  Jahnke K.,   H{\"a}u{\ss}ler B.,  2008, \mn@doi [ApJS]
  {10.1086/524039}, 175, 105

\bibitem[\protect\citeauthoryear{Bertin \& Arnouts}{Bertin \&
  Arnouts}{1996}]{Bertin96}
Bertin E.,  Arnouts S.,  1996, \mn@doi [A\&AS] {10.1051/aas:1996164}, 117, 393

\bibitem[\protect\citeauthoryear{Boada et~al.,}{Boada et~al.}{2015}]{Boada15}
Boada S.,  et~al., 2015, \mn@doi [ApJ] {10.1088/0004-637X/803/2/104}, 803, 104

\bibitem[\protect\citeauthoryear{Brammer et~al.,}{Brammer
  et~al.}{2012}]{Brammer12}
Brammer G.~B.,  et~al., 2012, \mn@doi [ApJS] {10.1088/0067-0049/200/2/13}, 200,
  13

\bibitem[\protect\citeauthoryear{Bruzual \& Charlot}{Bruzual \&
  Charlot}{2003}]{Bruzual03}
Bruzual G.,  Charlot S.,  2003, \mn@doi [MNRAS]
  {10.1046/j.1365-8711.2003.06897.x}, 344, 1000

\bibitem[\protect\citeauthoryear{Calzetti, Kinney  \&
  Storchi-Bergmann}{Calzetti et~al.}{1994}]{Calzetti94}
Calzetti D.,  Kinney A.~L.,   Storchi-Bergmann T.,  1994, \mn@doi [ApJ]
  {10.1086/174346}, 429, 582

\bibitem[\protect\citeauthoryear{Calzetti, Armus, Bohlin, Kinney, Koornneef  \&
  Storchi-Bergmann}{Calzetti et~al.}{2000}]{Calzetti00}
Calzetti D.,  Armus L.,  Bohlin R.~C.,  Kinney A.~L.,  Koornneef J.,
  Storchi-Bergmann T.,  2000, \mn@doi [ApJ] {10.1086/308692}, 533, 682

\bibitem[\protect\citeauthoryear{Cappellari \& Copin}{Cappellari \&
  Copin}{2003}]{Cappellari03}
Cappellari M.,  Copin Y.,  2003, \mn@doi [MNRAS]
  {10.1046/j.1365-8711.2003.06541.x}, 342, 345

\bibitem[\protect\citeauthoryear{Cardelli, Clayton  \& Mathis}{Cardelli
  et~al.}{1989}]{Cardelli89}
Cardelli J.~A.,  Clayton G.~C.,   Mathis J.~S.,  1989, \mn@doi [ApJ]
  {10.1086/167900}, 345, 245

\bibitem[\protect\citeauthoryear{Chabrier}{Chabrier}{2003}]{Chabrier03}
Chabrier G.,  2003, \mn@doi [PASP] {10.1086/376392}, 115, 763

\bibitem[\protect\citeauthoryear{Chan et~al.,}{Chan et~al.}{2016}]{Chan16}
Chan J. C.~C.,  et~al., 2016, \mn@doi [MNRAS] {10.1093/mnras/stw502}, 458, 3181

\bibitem[\protect\citeauthoryear{Cibinel et~al.,}{Cibinel
  et~al.}{2015}]{Cibinel15}
Cibinel A.,  et~al., 2015, \mn@doi [ApJ] {10.1088/0004-637X/805/2/181}, 805,
  181

\bibitem[\protect\citeauthoryear{Coil et~al.,}{Coil et~al.}{2015}]{Coil15}
Coil A.~L.,  et~al., 2015, \mn@doi [ApJ] {10.1088/0004-637X/801/1/35}, 801, 35

\bibitem[\protect\citeauthoryear{Conselice}{Conselice}{2003}]{Conselice03}
Conselice C.~J.,  2003, \mn@doi [ApJS] {10.1086/375001}, 147, 1

\bibitem[\protect\citeauthoryear{Conselice}{Conselice}{2014}]{Conselice14}
Conselice C.~J.,  2014, \mn@doi [ARA\&A] {10.1146/annurev-astro-081913-040037},
  52, 291

\bibitem[\protect\citeauthoryear{Dickinson}{Dickinson}{2000}]{Dickinson00}
Dickinson M.,  2000, \mn@doi [Philosophical Transactions of the Royal Society
  A: Mathematical, Physical and Engineering Sciences] {10.1098/rsta.2000.0626},
  358, 2001

\bibitem[\protect\citeauthoryear{Draine \& Lee}{Draine \& Lee}{1984}]{Draine84}
Draine B.~T.,  Lee H.~M.,  1984, \mn@doi [ApJ] {10.1086/162480}, 285, 89

\bibitem[\protect\citeauthoryear{Fitzpatrick \& Massa}{Fitzpatrick \&
  Massa}{1986}]{Fitzpatrick86}
Fitzpatrick E.~L.,  Massa D.,  1986, \mn@doi [ApJ] {10.1086/164415}, 307, 286

\bibitem[\protect\citeauthoryear{Fitzpatrick \& Massa}{Fitzpatrick \&
  Massa}{1990}]{Fitzpatrick90}
Fitzpatrick E.~L.,  Massa D.,  1990, \mn@doi [ApJS] {10.1086/191413}, 72, 163

\bibitem[\protect\citeauthoryear{Genzel et~al.,}{Genzel
  et~al.}{2013}]{Genzel13}
Genzel R.,  et~al., 2013, \mn@doi [ApJ] {10.1088/0004-637X/773/1/68}, 773, 68

\bibitem[\protect\citeauthoryear{Gordon, Clayton, Misselt, Landolt  \&
  Wolff}{Gordon et~al.}{2003}]{Gordon03}
Gordon K.~D.,  Clayton G.~C.,  Misselt K.~A.,  Landolt A.~U.,   Wolff M.~J.,
  2003, \mn@doi [ApJ] {10.1086/376774}, 594, 279

\bibitem[\protect\citeauthoryear{Griffiths et~al.,}{Griffiths
  et~al.}{1994}]{Griffiths94}
Griffiths R.~E.,  et~al., 1994, \mn@doi [ApJ] {10.1086/187584}, 435, L19

\bibitem[\protect\citeauthoryear{Grogin et~al.,}{Grogin
  et~al.}{2011}]{Grogin11}
Grogin N.~A.,  et~al., 2011, \mn@doi [ApJS] {10.1088/0067-0049/197/2/35}, 197,
  35

\bibitem[\protect\citeauthoryear{Guo, Giavalisco, Ferguson, Cassata  \&
  Koekemoer}{Guo et~al.}{2012}]{Guo12-1}
Guo Y.,  Giavalisco M.,  Ferguson H.~C.,  Cassata P.,   Koekemoer A.~M.,  2012,
  \mn@doi [ApJ] {10.1088/0004-637X/757/2/120}, 757, 120

\bibitem[\protect\citeauthoryear{Guo et~al.,}{Guo et~al.}{2015}]{Guo15}
Guo Y.,  et~al., 2015, \mn@doi [ApJ] {10.1088/0004-637X/800/1/39}, 800, 39

\bibitem[\protect\citeauthoryear{Guo et~al.,}{Guo et~al.}{2018}]{Guo18}
Guo Y.,  et~al., 2018, \mn@doi [ApJ] {10.3847/1538-4357/aaa018}, 853, 108

\bibitem[\protect\citeauthoryear{Hemmati et~al.,}{Hemmati
  et~al.}{2014}]{Hemmati14}
Hemmati S.,  et~al., 2014, \mn@doi [ApJ] {10.1088/0004-637X/797/2/108}, 797,
  108

\bibitem[\protect\citeauthoryear{Hubble}{Hubble}{1926}]{Hubble26}
Hubble E.~P.,  1926, \mn@doi [ApJ] {10.1086/143018}, 64

\bibitem[\protect\citeauthoryear{Hunter}{Hunter}{2007}]{Hunter07}
Hunter J.~D.,  2007, \mn@doi [Computing in Science \& Engineering]
  {10.1109/MCSE.2007.55}, 9, 90

\bibitem[\protect\citeauthoryear{Jung et~al.,}{Jung et~al.}{2017}]{Jung17}
Jung I.,  et~al., 2017, \mn@doi [ApJ] {10.3847/1538-4357/834/1/81}, 834, 81

\bibitem[\protect\citeauthoryear{Koekemoer et~al.,}{Koekemoer
  et~al.}{2011}]{Koekemoer11}
Koekemoer A.~M.,  et~al., 2011, \mn@doi [ApJS] {10.1088/0067-0049/197/2/36},
  197, 36

\bibitem[\protect\citeauthoryear{Kriek et~al.,}{Kriek et~al.}{2015}]{Kriek15}
Kriek M.,  et~al., 2015, \mn@doi [ApJS] {10.1088/0067-0049/218/2/15}, 218, 15

\bibitem[\protect\citeauthoryear{Lang et~al.,}{Lang et~al.}{2014}]{Lang14}
Lang P.,  et~al., 2014, \mn@doi [ApJ] {10.1088/0004-637X/788/1/11}, 788, 11

\bibitem[\protect\citeauthoryear{Leung et~al.,}{Leung et~al.}{2019}]{Leung19}
Leung G. C.~K.,  et~al., 2019, \mn@doi [ApJ] {10.3847/1538-4357/ab4a7c}, 886,
  11

\bibitem[\protect\citeauthoryear{Lotz, Primack  \& Madau}{Lotz
  et~al.}{2004}]{Lotz04}
Lotz J.~M.,  Primack J.,   Madau P.,  2004, \mn@doi [AJ] {10.1086/421849}, 128,
  163

\bibitem[\protect\citeauthoryear{Lotz et~al.,}{Lotz et~al.}{2008}]{Lotz08}
Lotz J.~M.,  et~al., 2008, \mn@doi [ApJ] {10.1086/523659}, 672, 177

\bibitem[\protect\citeauthoryear{Madau \& Dickinson}{Madau \&
  Dickinson}{2014}]{Madau14}
Madau P.,  Dickinson M.,  2014, \mn@doi [ARA\&A]
  {10.1146/annurev-astro-081811-125615}, 52, 415

\bibitem[\protect\citeauthoryear{McLean et~al.,}{McLean
  et~al.}{2010}]{McLean10}
McLean I.~S.,  et~al., 2010, in Ground-based and {Airborne} {Instrumentation}
  for {Astronomy} {III}. Proceedings of the SPIE, p. 77351E,
  \mn@doi{10.1117/12.856715}, \url
  {http://adsabs.harvard.edu/abs/2010SPIE.7735E..1EM}

\bibitem[\protect\citeauthoryear{McLean et~al.,}{McLean
  et~al.}{2012}]{McLean12}
McLean I.~S.,  et~al., 2012, in Ground-based and {Airborne} {Instrumentation}
  for {Astronomy} {IV}. p. 84460J, \mn@doi{10.1117/12.924794}, \url
  {http://adsabs.harvard.edu/abs/2012SPIE.8446E..0JM}

\bibitem[\protect\citeauthoryear{Momcheva et~al.,}{Momcheva
  et~al.}{2016}]{Momcheva16}
Momcheva I.~G.,  et~al., 2016, \mn@doi [ApJS] {10.3847/0067-0049/225/2/27},
  225, 27

\bibitem[\protect\citeauthoryear{Oke \& Gunn}{Oke \& Gunn}{1983}]{Oke83}
Oke J.~B.,  Gunn J.~E.,  1983, \mn@doi [ApJ] {10.1086/160817}, 266, 713

\bibitem[\protect\citeauthoryear{Oliphant}{Oliphant}{2007}]{Oliphant07}
Oliphant T.~E.,  2007, \mn@doi [Computing in Science \& Engineering]
  {10.1109/MCSE.2007.58}, 9, 10

\bibitem[\protect\citeauthoryear{Papovich, Dickinson, Giavalisco, Conselice  \&
  Ferguson}{Papovich et~al.}{2005}]{Papovich05}
Papovich C.,  Dickinson M.,  Giavalisco M.,  Conselice C.~J.,   Ferguson H.~C.,
   2005, \mn@doi [ApJ] {10.1086/429120}, 631, 101

\bibitem[\protect\citeauthoryear{Reddy, Pettini, Steidel, Shapley, Erb  \&
  Law}{Reddy et~al.}{2012}]{Reddy12-1}
Reddy N.~A.,  Pettini M.,  Steidel C.~C.,  Shapley A.~E.,  Erb D.~K.,   Law
  D.~R.,  2012, \mn@doi [ApJ] {10.1088/0004-637X/754/1/25}, 754, 25

\bibitem[\protect\citeauthoryear{Reddy et~al.,}{Reddy et~al.}{2015}]{Reddy15}
Reddy N.~A.,  et~al., 2015, \mn@doi [ApJ] {10.1088/0004-637X/806/2/259}, 806,
  259

\bibitem[\protect\citeauthoryear{Reddy et~al.,}{Reddy et~al.}{2018}]{Reddy18-1}
Reddy N.~A.,  et~al., 2018, \mn@doi [ApJ] {10.3847/1538-4357/aaa3e7}, 853, 56

\bibitem[\protect\citeauthoryear{Scott et~al.,}{Scott et~al.}{2017}]{Scott17}
Scott N.,  et~al., 2017, \mn@doi [MNRAS] {10.1093/mnras/stx2166}, 472, 2833

\bibitem[\protect\citeauthoryear{S{\'e}rsic}{S{\'e}rsic}{1963}]{Sersic63}
S{\'e}rsic J.,  1963, BAAA, 6, 41

\bibitem[\protect\citeauthoryear{Shapley}{Shapley}{2011}]{Shapley11}
Shapley A.~E.,  2011, \mn@doi [ARA\&A] {10.1146/annurev-astro-081710-102542},
  49, 525

\bibitem[\protect\citeauthoryear{Shapley, Steidel, Adelberger, Dickinson,
  Giavalisco  \& Pettini}{Shapley et~al.}{2001}]{Shapley01}
Shapley A.~E.,  Steidel C.~C.,  Adelberger K.~L.,  Dickinson M.,  Giavalisco
  M.,   Pettini M.,  2001, \mn@doi [ApJ] {10.1086/323432}, 562, 95

\bibitem[\protect\citeauthoryear{Shivaei et~al.,}{Shivaei
  et~al.}{2015}]{Shivaei15}
Shivaei I.,  et~al., 2015, \mn@doi [ApJ] {10.1088/0004-637X/815/2/98}, 815, 98

\bibitem[\protect\citeauthoryear{Shivaei et~al.,}{Shivaei
  et~al.}{2020}]{Shivaei20}
Shivaei I.,  et~al., 2020, preprint, 2005, arXiv:2005.01742

\bibitem[\protect\citeauthoryear{Skelton et~al.,}{Skelton
  et~al.}{2014}]{Skelton14}
Skelton R.~E.,  et~al., 2014, \mn@doi [ApJS] {10.1088/0067-0049/214/2/24}, 214,
  24

\bibitem[\protect\citeauthoryear{Tacchella et~al.,}{Tacchella
  et~al.}{2018}]{Tacchella18}
Tacchella S.,  et~al., 2018, \mn@doi [ApJ] {10.3847/1538-4357/aabf8b}, 859, 56

\bibitem[\protect\citeauthoryear{Tadaki, Kodama, Tanaka, Hayashi, Koyama  \&
  Shimakawa}{Tadaki et~al.}{2014}]{Tadaki14}
Tadaki K.-i.,  Kodama T.,  Tanaka I.,  Hayashi M.,  Koyama Y.,   Shimakawa R.,
  2014, \mn@doi [ApJ] {10.1088/0004-637X/780/1/77}, 780, 77

\bibitem[\protect\citeauthoryear{Toft et~al.,}{Toft et~al.}{2007}]{Toft07}
Toft S.,  et~al., 2007, \mn@doi [ApJ] {10.1086/521810}, 671, 285

\bibitem[\protect\citeauthoryear{Torrey et~al.,}{Torrey
  et~al.}{2015}]{Torrey15}
Torrey P.,  et~al., 2015, \mn@doi [MNRAS] {10.1093/mnras/stu2592}, 447, 2753

\bibitem[\protect\citeauthoryear{Trujillo, Conselice, Bundy, Cooper, Eisenhardt
   \& Ellis}{Trujillo et~al.}{2007}]{Trujillo07}
Trujillo I.,  Conselice C.~J.,  Bundy K.,  Cooper M.~C.,  Eisenhardt P.,
  Ellis R.~S.,  2007, \mn@doi [MNRAS] {10.1111/j.1365-2966.2007.12388.x}, 382,
  109

\bibitem[\protect\citeauthoryear{Worthey}{Worthey}{1994}]{Worthey94}
Worthey G.,  1994, \mn@doi [ApJS] {10.1086/192096}, 95, 107

\bibitem[\protect\citeauthoryear{Wuyts et~al.,}{Wuyts et~al.}{2012}]{Wuyts12}
Wuyts S.,  et~al., 2012, \mn@doi [ApJ] {10.1088/0004-637X/753/2/114}, 753, 114

\bibitem[\protect\citeauthoryear{Wuyts et~al.,}{Wuyts et~al.}{2013}]{Wuyts13}
Wuyts S.,  et~al., 2013, \mn@doi [ApJ] {10.1088/0004-637X/779/2/135}, 779, 135

\bibitem[\protect\citeauthoryear{Wuyts, Rigby, Gladders  \& Sharon}{Wuyts
  et~al.}{2014}]{Wuyts14}
Wuyts E.,  Rigby J.~R.,  Gladders M.~D.,   Sharon K.,  2014, \mn@doi [ApJ]
  {10.1088/0004-637X/781/2/61}, 781, 61

\bibitem[\protect\citeauthoryear{van~de Sande et~al.,}{van~de Sande
  et~al.}{2018}]{van_de_Sande18}
van~de Sande J.,  et~al., 2018, \mn@doi [Nature Astronomy]
  {10.1038/s41550-018-0436-x}, 2, 483

\bibitem[\protect\citeauthoryear{van~der Wel et~al.,}{van~der Wel
  et~al.}{2014}]{van_Der_Wel14}
van~der Wel A.,  et~al., 2014, \mn@doi [ApJ] {10.1088/0004-637X/788/1/28}, 788,
  28

\makeatother
\end{thebibliography}

\appendix

\section{The Wuyts et al. (2012) Method}\label{app:wuyts}
\begin{figure*}
\includegraphics[width=.9\textwidth]{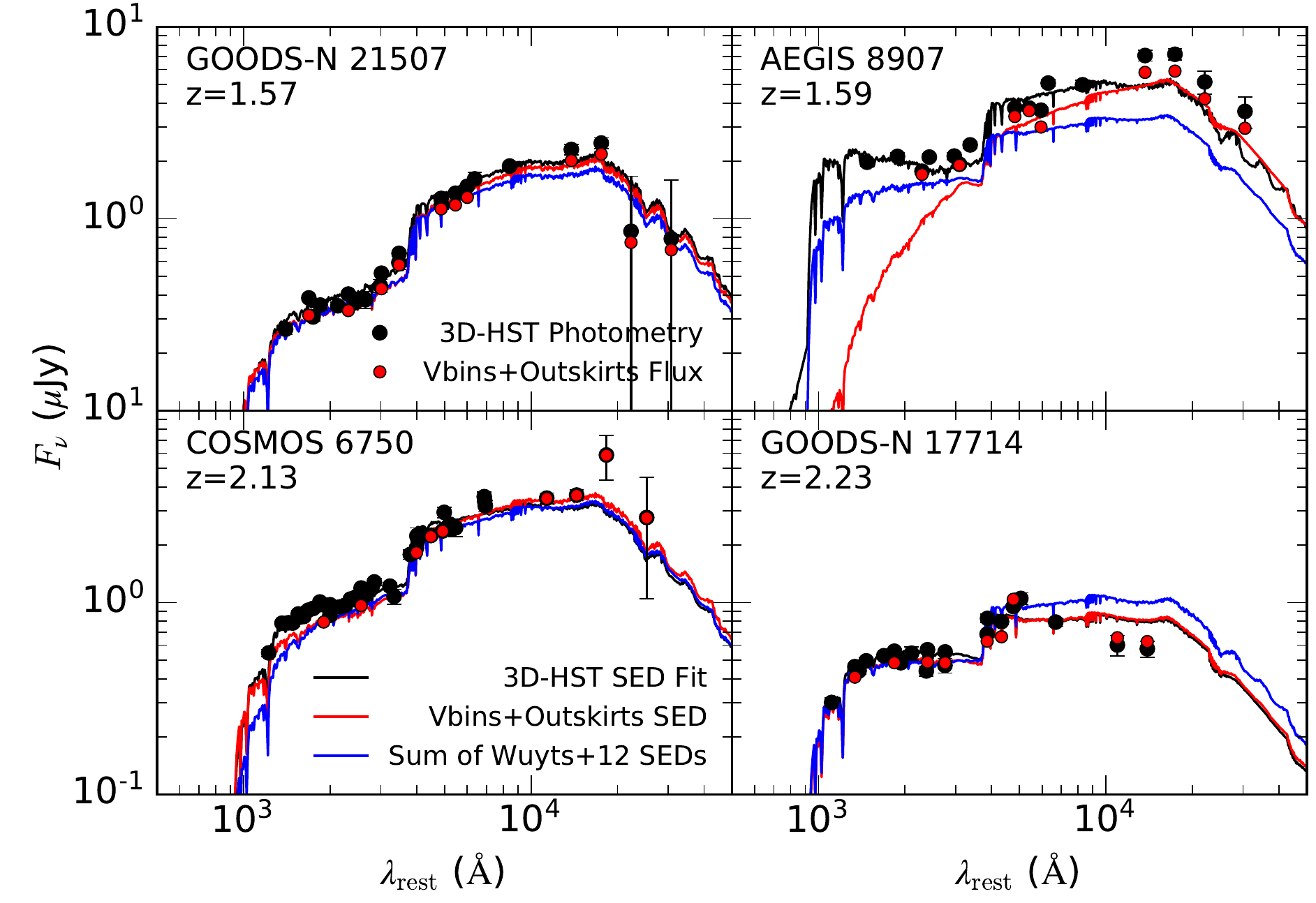}
\caption{Best-fit integrated SEDs from three different SED fitting methods for four example galaxies. The colors of the points and curves are the same as \autoref{fig:sedcomp}. The black points and curves shows the fluxes from the 3D-HST photometry and the respective best-fit SED. The red points and curves indicate the total flux from all Voronoi bins (including outskirts) and the sum of their best-fit resolved SEDs for the IRAC data that has been normalized to the $H_{160}$ flux (see \autoref{sec:flux} and \autoref{eq:irac}). The dark blue curves show the sum of the resolved SEDs that have alternatively been modified to include the IRAC photometry using the method described by \citet{Wuyts12}. The errors are typically smaller than the size of the points. Note that not all galaxies have outskirt components, such as GOODS-N\,21507.}
\label{fig:intspec_wuyts}
\end{figure*}
\begin{figure*}
\includegraphics[width=.8\textwidth]{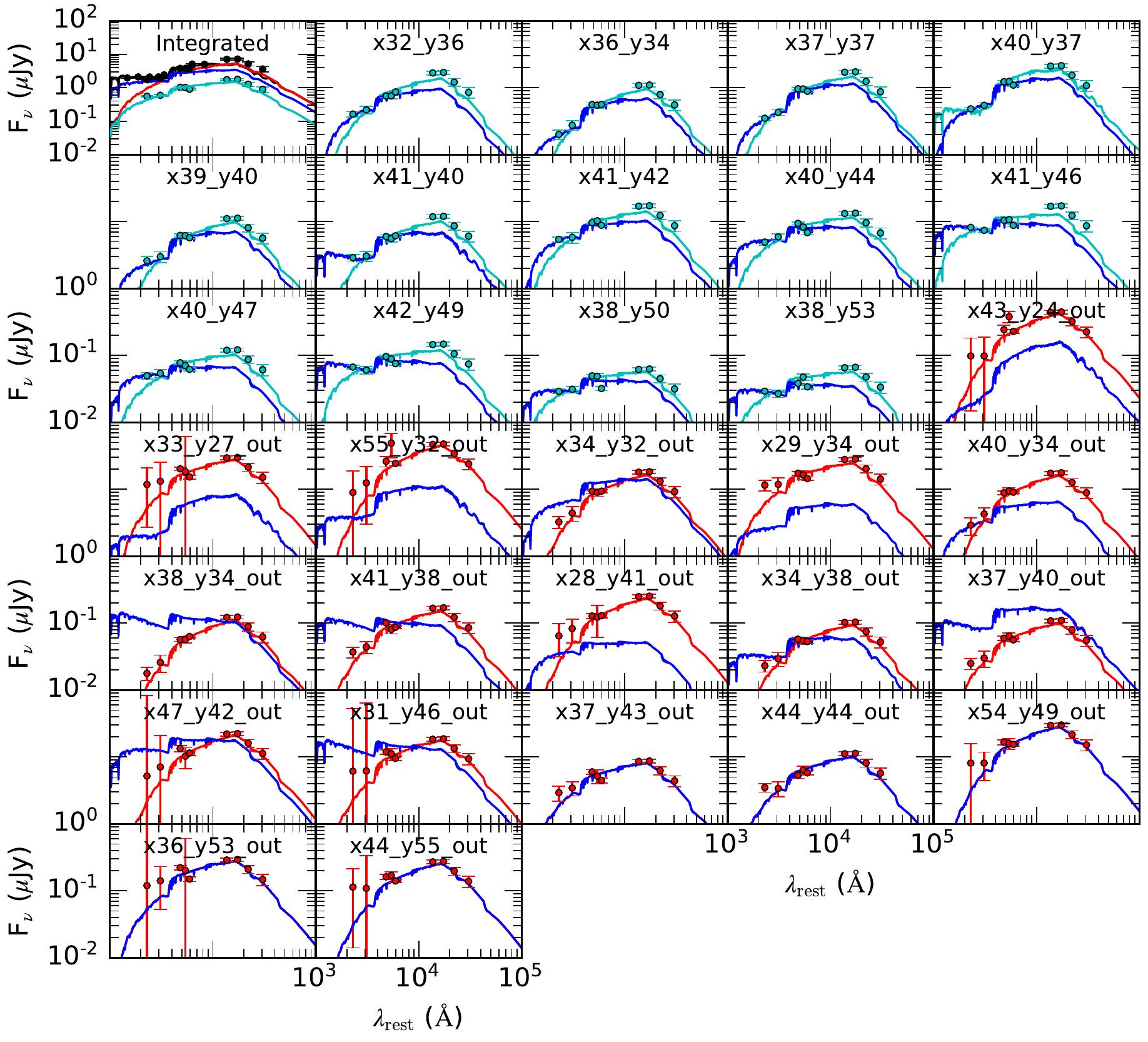}
\includegraphics[width=.8\textwidth]{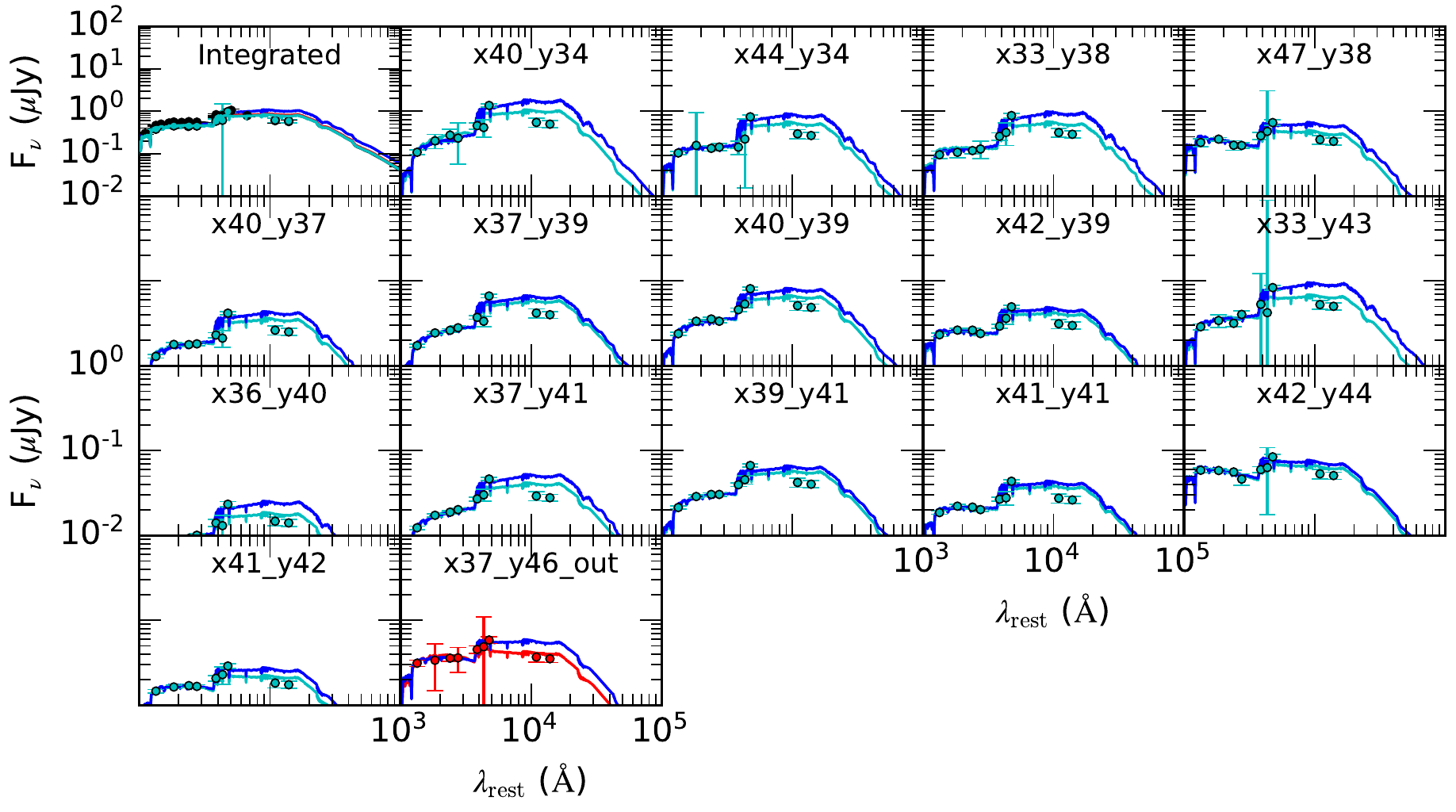}
\caption{The resolved SEDs for every Voronoi and outskirt bin comparing the two methods that incorporate the unresolved IRAC photometry for AEGIS\,8907 (\textit{top panels}) and GOODS-N\,17714 (\textit{bottom panels}). The colors of the points and curves are the same as \autoref{fig:sedcomp} and \autoref{fig:intspec_wuyts}. The first panel is the same as \autoref{fig:sedcomp}, but also includes the light blue points and curves that show the sum of the flux and SEDs from the well-constrained Voronoi bins (S/N $\geq5$ in at least 5\,filters). The panels with light blue points and curves show resolved SEDs fit to Voronoi bins and the panels with red points and curves show resolved SEDs fit to outskirt bins. The light blue and red SED curves directly include the IRAC photometry in the resolved SED fitting by normalizing the IRAC flux by the $H_{160}$ resolved flux (see \autoref{eq:irac}). The dark blue curves show the resolved SEDs that have been modified to include the contribution of the IRAC photometry using the method described by \citet{Wuyts12}. Note that in some cases, particularly in the outskirt bins, there is no change between the original SED and the SED modified by the Wuyts et al. method, such that the SED curves overlap in the figure. The labels in each panel indicate the central pixel coordinate of each bin.}
\label{fig:allspec_ex}
\end{figure*}
\begin{figure*}
\begin{adjustbox}{width=\linewidth, center}
\begin{minipage}{\linewidth}
\includegraphics[width=\linewidth]{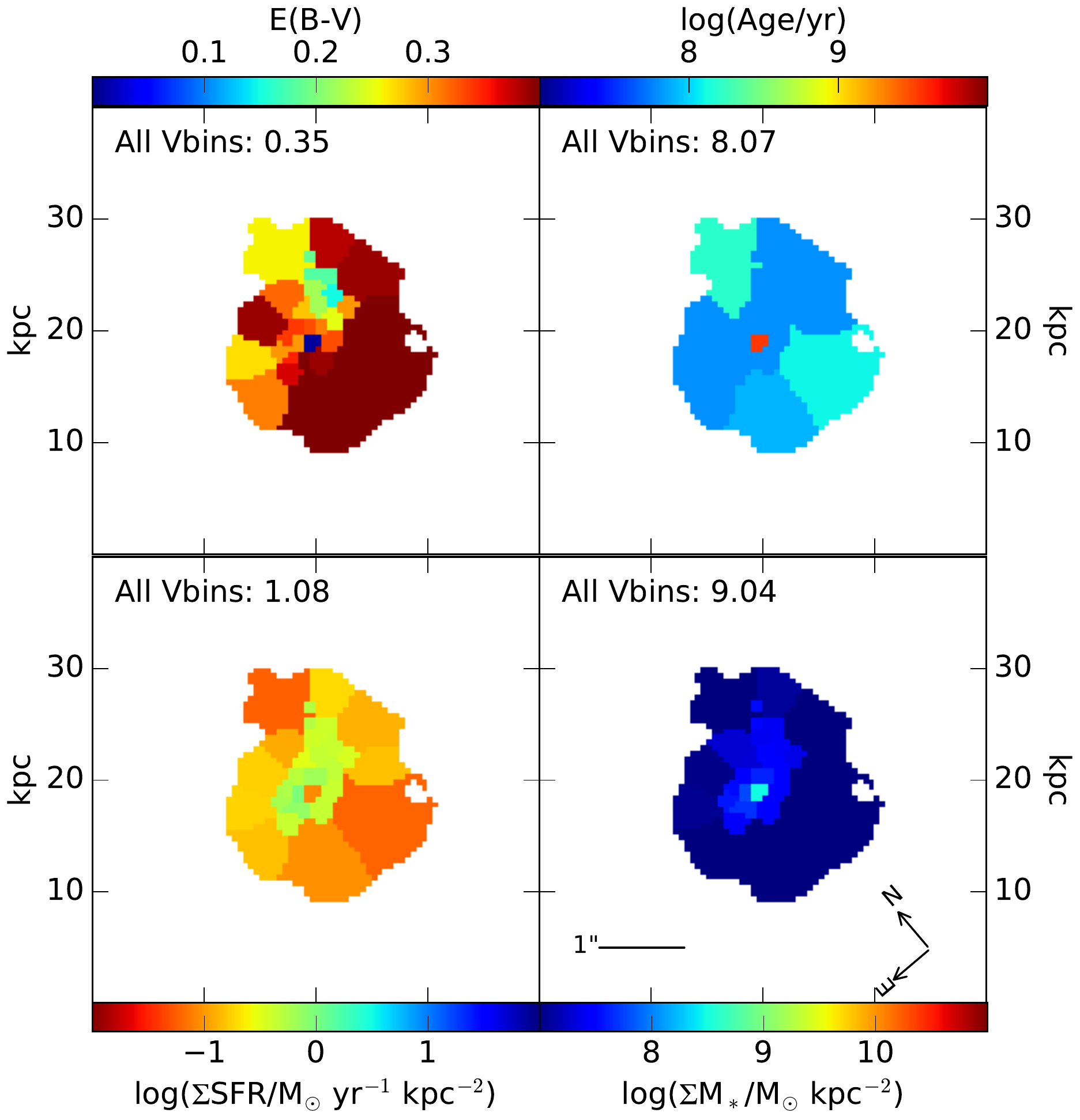}
\end{minipage}
\quad
\begin{minipage}{\linewidth}
\includegraphics[width=\linewidth]{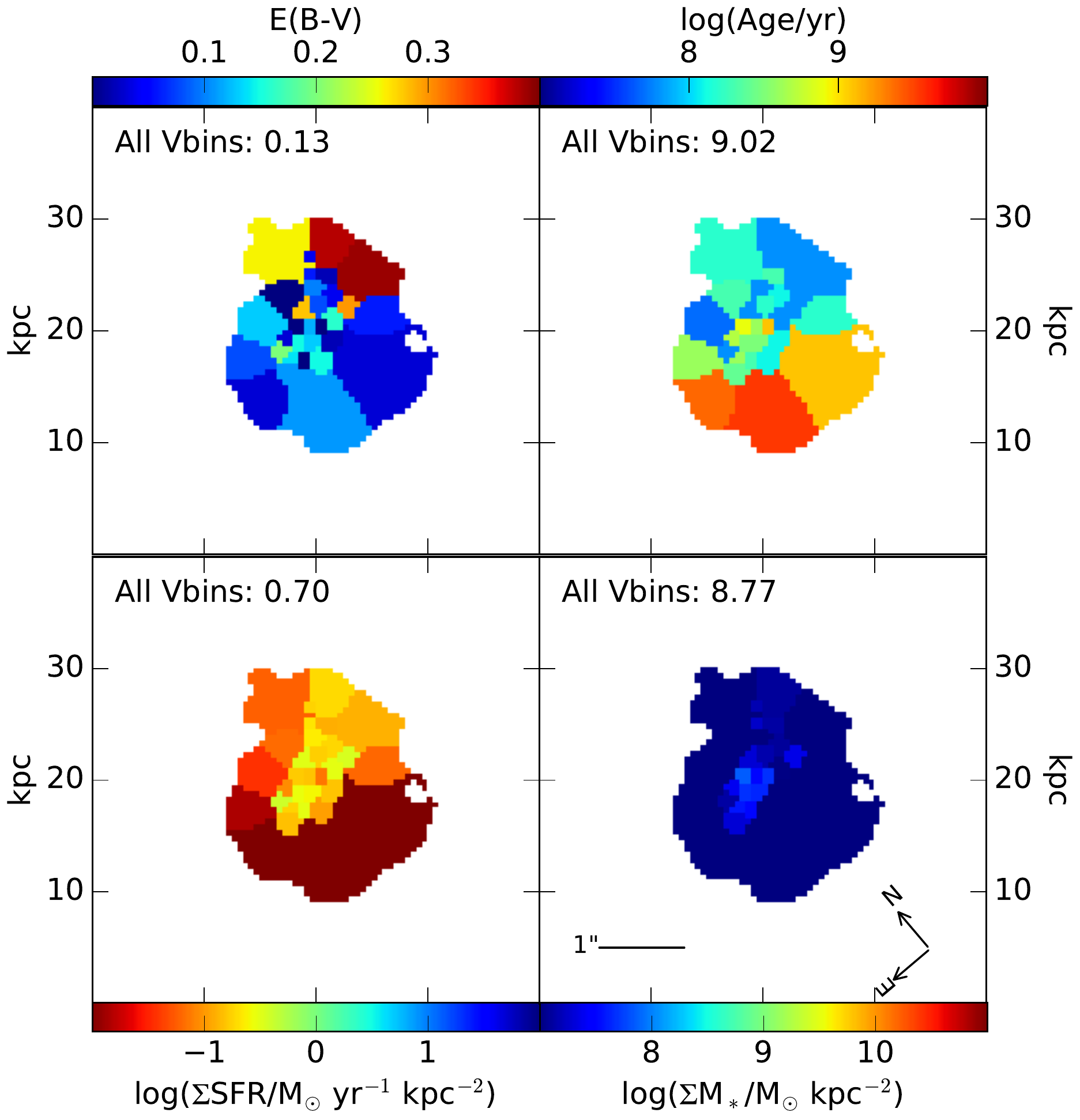}
\end{minipage}
\end{adjustbox}
\begin{adjustbox}{width=\linewidth, center}
\begin{minipage}{\linewidth}
\includegraphics[width=\linewidth]{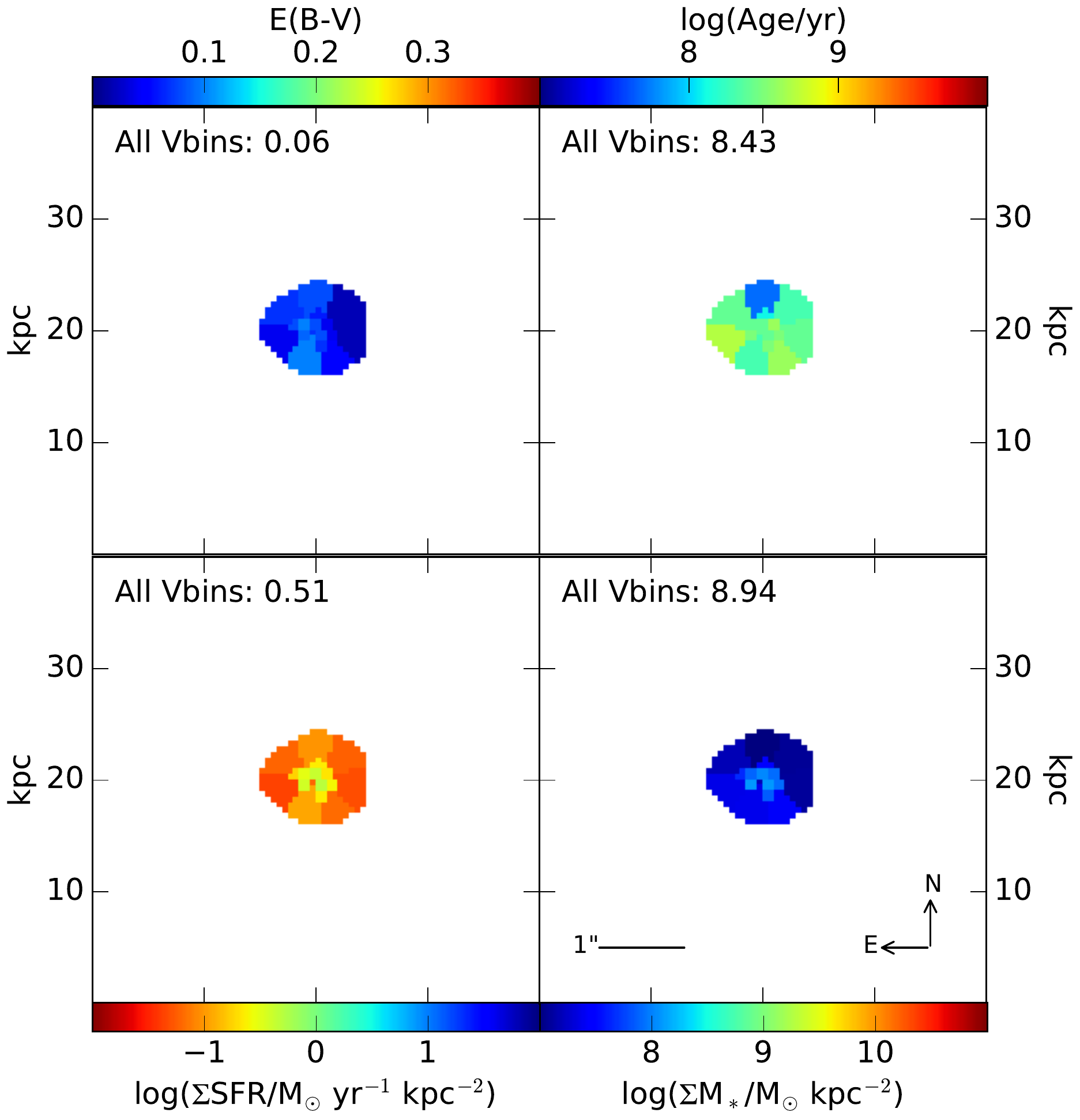}
\end{minipage}
\quad
\begin{minipage}{\linewidth}
\includegraphics[width=\linewidth]{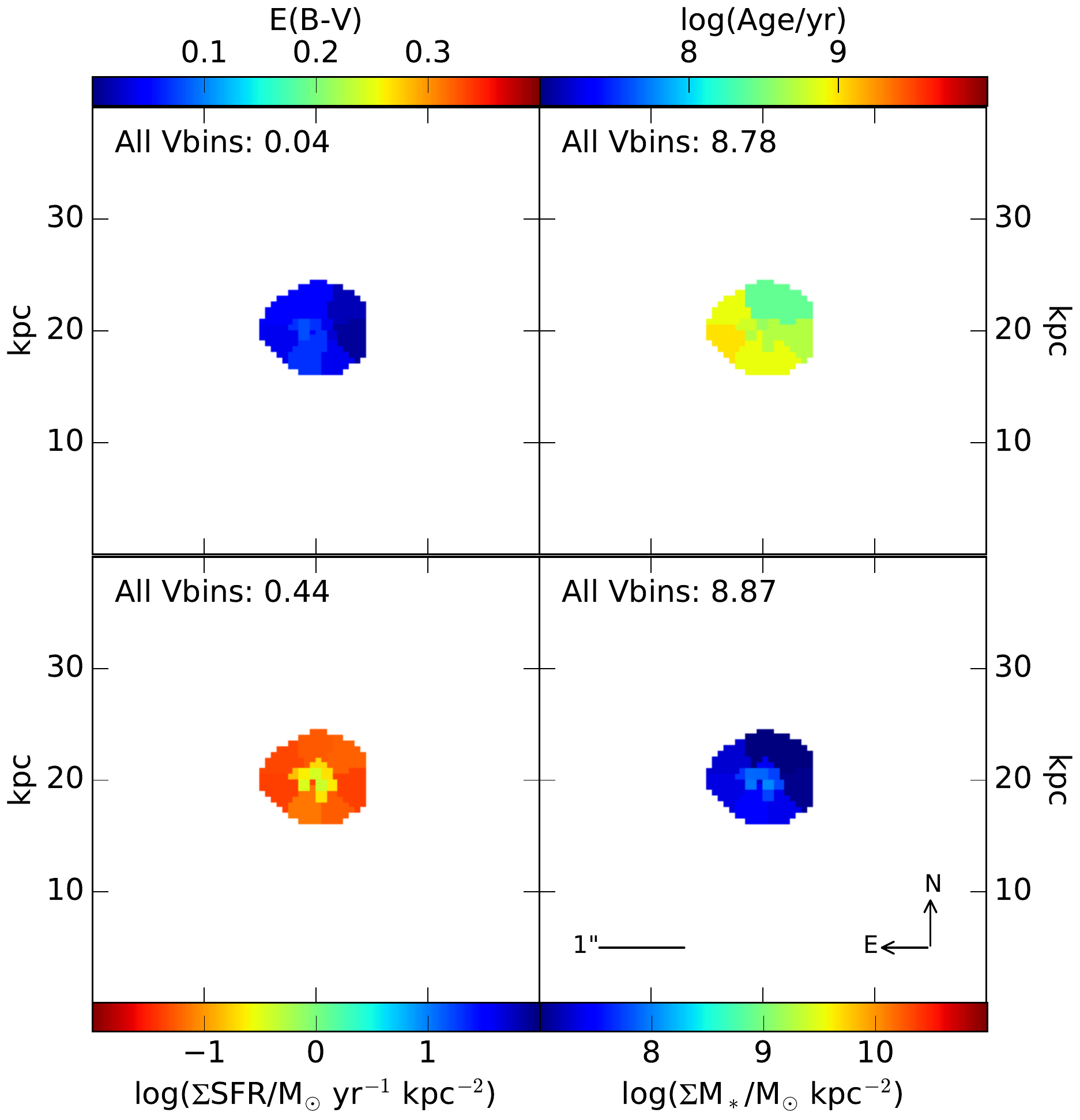}
\end{minipage}
\end{adjustbox}
\caption{Stellar population and reddening maps from the resolved SED fitting (including outskirt components) that incorporates IRAC photometry using a normalization based on the $H_{160}$ flux (\textit{left panels}) and the method described by \citet{Wuyts12} (\textit{right panels}). The distribution of \EBVs\ (\textit{top left}), stellar population age (\textit{top right}), SFR (\textit{bottom left}), and stellar mass (\textit{bottom right}) are shown for AEGIS\,8907 ($z=1.59$, \textit{top panels}) and GOODS-N\,17714 ($z=2.23$, \textit{bottom panels}). The listed number in each panel shows the average \EBVs\ or stellar population age weighted by the Voronoi bin areas or the summed SFR or stellar mass from all of the Voronoi bins.}
\label{fig:map_ex_wuyts}
\end{figure*}
In order to best constrain stellar mass estimates determined from the resolved SED fitting and constrain the shape of the resolved SEDs in the near-IR, we elect to include the unresolved \textit{Spitzer}/IRAC photometry with the resolved photometry. \citet{Wuyts12} presented a method for incorporating unresolved photometry into the results of resolved SED fitting by modifying the best-fit resolved SEDs such that their sum produces an SED that is optimally fit to the integrated photometry. We alternatively use a simple normalization based on the $H_{160}$ flux to directly use the IRAC photometry in the resolved SED fitting (see \autoref{sec:flux} and \autoref{eq:irac}). In this appendix we compare the best-fit SEDs and SED-derived properties that result from the simple normalization we assumed versus the Wuyts et al. methodology. For this investigation we select GOODS-N\,21507 ($z=1.57$), AEGIS\,8907 ($z=1.59$), COSMOS\,6750 ($z=2.13$), and GOODS-N\,17714 ($z=2.23$), thus including two galaxies from our $z\sim1.5$ and $z\sim2.3$ redshift bins.

Here we briefly describe the method, but refer the reader to \citet{Wuyts12} for a detailed description. First an initial fit is obtained for all resolved components using data from the resolved \HST\ imaging by finding the model where the resolved chi-squared, $\chi_\text{res}$, is minimized (see \autoref{sec:sedfit_res}). The sum of the resolved SEDs is then compared to the integrated photometry---including the desired unresolved filters---by considering the integrated chi-squared defined by
\begin{equation}
\chi^2_{\text{res+int}} = \chi^2_{\text{res}} + \sum^{N_{\text{filt}}}_{j=1} \frac{(F_j - \Sigma^{N_{\text{Vbin}}}_{i=1} M_{i,j})^2}{\sigma_j^2}
\text{,}
\end{equation}
where $F_j$ and $\sigma_j$ are the integrated flux and flux errors in each $j$ unresolved filter, respectively, and $M_{i,j}$ is the flux of the corresponding stellar population model in each $i$ Voronoi bin and $j$ unresolved filter for the $N_\text{filt}$ unresolved filters to be included \citep[in this case the \textit{Spitzer}/IRAC photometry; see Equation 2 from][]{Wuyts12}. The integrated chi-squared is iteratively improved by considering the next-best stellar population model in all resolved bins (worse resolved chi-squared) and selecting the bin that results in the largest change in the integrated chi-squared. This process is repeated up to 500 times or until there is no further improvement to be made in the integrated chi-squared. 

In \autoref{fig:intspec_wuyts}, we show the sum of the resolved SEDs from both our method (red curves) and the Wuyts et al. method (dark blue curves) compared to the SED that is fit to the unresolved 3D-HST photometry (black curves and points). The contribution from outskirt bins are included to best compare the summed SEDs with the SED that is fit to the unresolved 3D-HST photometry, but we note that a small normalization offset remains in some cases due to minor differences in the apertures (see the beginning of \autoref{sec:integrated}).\footnote{To correct for this offset, \citet{Wuyts12} additionally normalized the integrated SED to the 3D-HST $H_{160}$ flux. We do not include this additional normalization here since the purpose of this appendix is to compare our methodology with Wuyts et al. rather than reproduce the 3D-HST photometry. The change in normalization causes an insignificant change in the total SFR and stellar mass and does not affect the results presented here.} With the exception of AEGIS\,8907, the integrated SEDs are consistent with each other. As discussed in \autoref{sec:intparam}, the UV slope of AEGIS\,8907 is not well constrained due to there only being one resolved filter available at 1250--2500\,\AA. Therefore, our SED fitting procedure prioritizes fitting the data in the near-IR that is covered by the IRAC photometry, whereas the Wuyts et al. method emphasizes the shape of the optical photometry initially and later adjusts to best-match the integrated 3D-HST SED. 

The resolved SEDs that are fit to the flux contained within each Voronoi bin are compared between our method and the Wuyts et al. method for AEGIS\,8907 and GOODS-N\,17714 in \autoref{fig:allspec_ex}. We remind the reader that the SEDs from the Wuyts et al. method (dark blue curves) show the SEDs that have been adjusted to produce a better match to the integrated photometry, such that the stellar population model has changed and the individual resolved SEDs may not necessarily match the resolved UV and optical photometry. \autoref{fig:allspec_ex} shows that the resolved SEDs are most discrepant in outskirt bins (panels with red points and curves), whereas the resolved SEDs generally agree between the two methods for the well-constrained Voronoi bins (S/N $\geq5$ in at least 5\,filters; panels with light blue points and curves). It is important to note that because the resolved stellar population models have changed in the Wuyts et al. method, the distribution of resolved properties may not be well-preserved---particularly in outskirt bins where the resolved SEDs are less constrained. However, in cases such as AEGIS\,8907, the Wuyts et al. method better reproduces the overall shape of the integrated 3D-HST photometry since the UV slope is unconstrained by the resolved photometry for low-redshift objects using our methodology (see \autoref{sec:intparam}). In most cases though, the sum of all the resolved SEDs from both methods represent good fits to the unresolved photometry. Finally, the resolved stellar population and dust maps (including outskirts) are shown in \autoref{fig:map_ex_wuyts} for AEGIS\,8907 (top panels) and GOODS-N\,17714 (bottom panels) from our method (left panels) and the Wuyts et al. method (right panels). We find that most cases are similar to GOODS-N\,17714, where the resolved SED properties are comparable between the two methods. For cases like AEGIS\,8907, the SFR and stellar mass distributions are also consistent between the two methods, with the greatest deviation being in the \EBVs\ distribution. As stated previously, the UV slope of AEGIS\,8907 is not well-constrained by the resolved photometry alone, such that the discrepancy in the resolved \EBVs\ distribution is unsurprising. 

In this work, we are interested in best preserving the distribution of the resolved stellar populations. Therefore, we elect to directly include the IRAC photometry in the resolved SED fitting by normalizing the IRAC photometry by the $H_{160}$ flux contained in each resolved element (see \autoref{eq:irac}). In this appendix, we have shown that the results obtained through the Wuyts et al. method are comparable to those obtained through our methodology. However, the Wuyts et al. method may be more appropriate in situations where key SED features (e.g., UV slope, Balmer/4000\,\AA\ breaks) cannot be constrained by the resolved photometry, but at the cost of the specific distribution of stellar population properties being more uncertain. Throughout our analysis, we retain separation between our $z\sim1.5$ and $z\sim2.3$ samples due to the discrepancies in the inferred \EBVs\ caused by the UV slope being unconstrained by the resolved photometry. 

\section{\HIRAC\ Color Perturbations}\label{app:colors}
\begin{figure*}
\includegraphics[width=\textwidth]{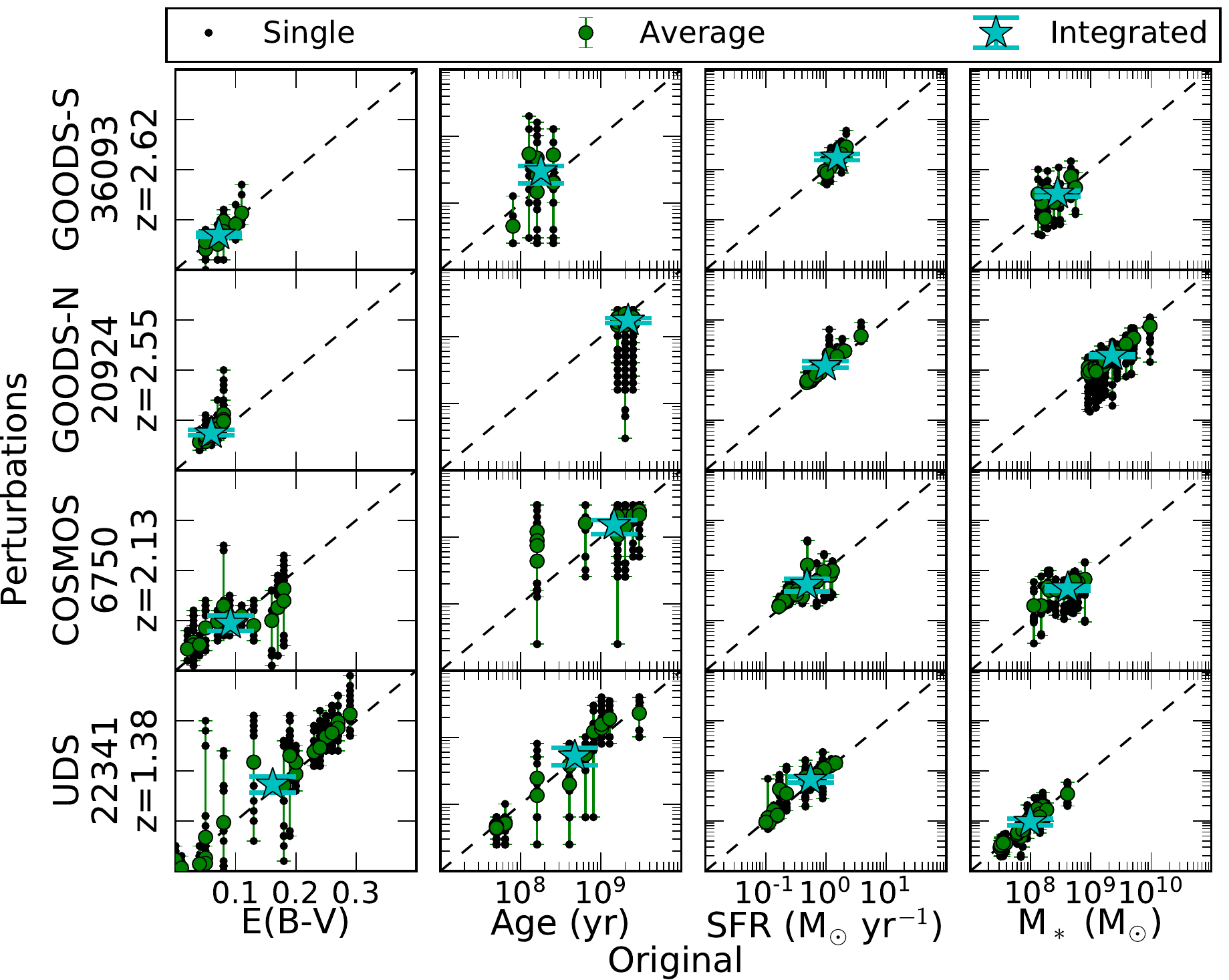}
\caption{The SED-derived \EBVs, stellar population ages, SFRs, and stellar masses for each Voronoi bin as the \HIRAC\ colors are perturbed. Each row represents a single galaxy from our sample for a spread of redshifts. Each vertically connected green line represents a single Voronoi bin that has been perturbed $n$ times, where $n$ is the number of Voronoi bins in the galaxy. The black points along a single vertical line show how the SED-derived properties change as that single Voronoi bin is perturbed, compared to the original values. Each green circle is the average of the perturbed results within a single Voronoi bin, compared to the original value. The light blue stars compare the average of the Voronoi bins over $n$ perturbations with the average of the original distribution (see \autoref{fig:perturb_ex}). The errors of the blue stars show the spread in the averaged Voronoi bins over all $n$ perturbations. The dashed black lines indicate where the perturbed properties equal the original properties.}
\label{fig:perturb_compare}
\end{figure*}
\begin{figure*}
\includegraphics[width=\textwidth]{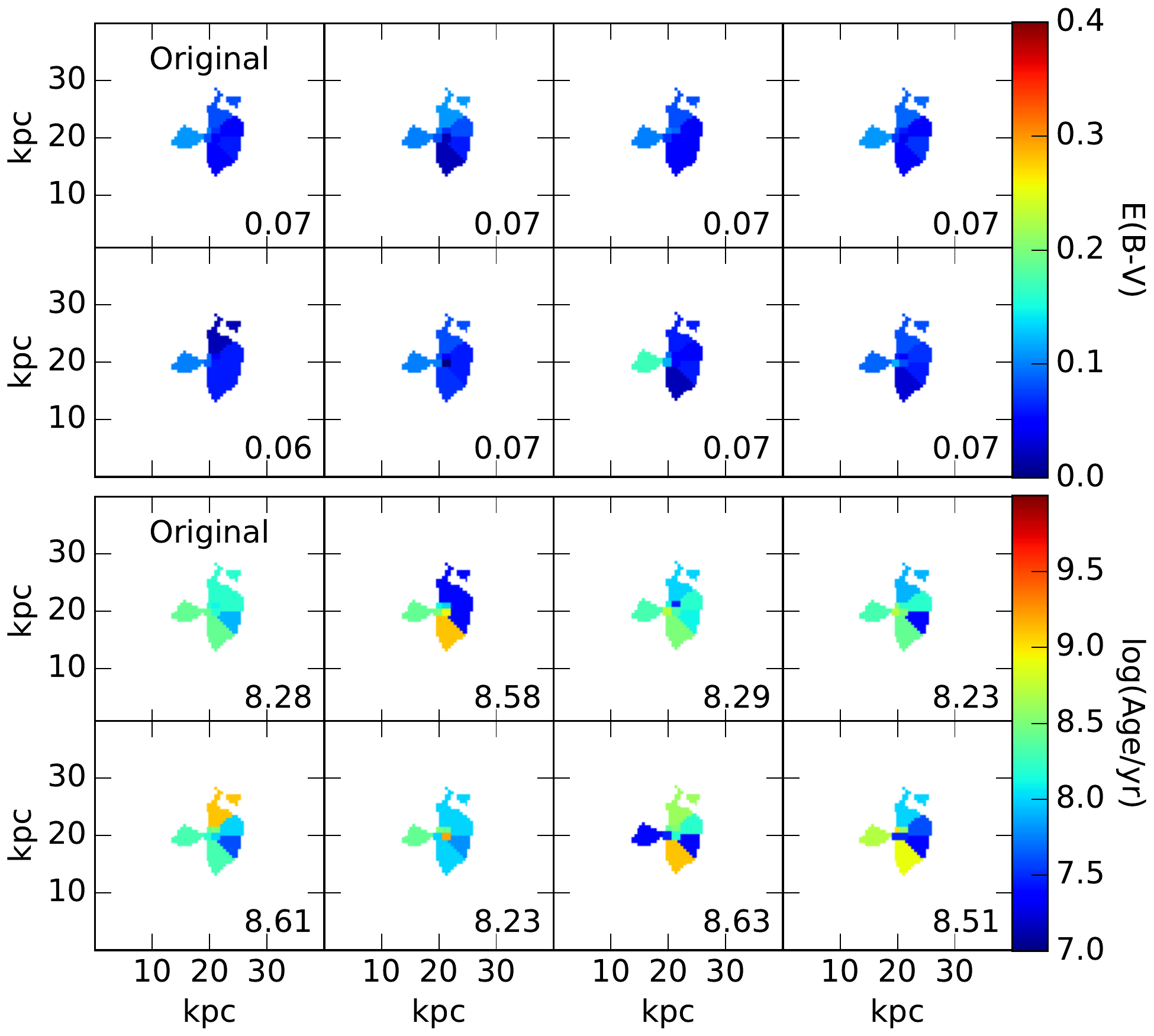}
\caption{\EBVs\ (\textit{top}) and stellar population age (\textit{bottom}) maps for GOODS-S\,36093 ($z=2.62$). The top left panels show the original \EBVs\ and stellar population age distribution from SED fitting, where the \HIRAC\ colors are constant across all Voronoi bins (see \autoref{eq:irac}). All other panels are show the \EBVs\ and stellar population age distributions when SED fitting was performed on the \HIRAC\ color perturbations. The value in the bottom right of each panel is the average \EBVs\ or log stellar population age (in dex) derived from the entire galaxy area.}
\label{fig:perturb_ex}
\label{lastpage}
\end{figure*}
In \autoref{sec:flux}, we normalized the unresolved {\it Spitzer}/IRAC photometry by the $H_{160}$ flux in each resolved element to accommodate for different area coverage. However, this unrealistically assumes that the IRAC flux directly traces the $H_{160}$-band flux. Most critically, this impacts galaxies at redshifts $z\gtrsim2.5$, where the Balmer/4000\,\AA~breaks shift into the $H_{160}$ filter and the break strength becomes constrained exclusively by the IRAC photometry. As a probe of stellar population age, the strength of the Balmer/4000\,\AA~breaks is crucial towards breaking the age-extinction degeneracy \citep[e.g.,][]{Worthey94, Shapley01}, where the shape of the SED in the rest-frame UV to near-IR of a young, dusty stellar population looks similar to an old, dust-free population. In this appendix the \HIRAC\ colors are perturbed to simulate spatially resolved variations in the IRAC photometry to assess how constraining the \HIRAC\ colors influences the integrated and resolved stellar population properties, with special care towards the stellar population ages and reddening. For this investigation we select GOODS-S\,36093, GOODS-N\,20924, COSMOS\,6750, and UDS\,22341, thus including two galaxies with redshifts above (2.62 and 2.55) and below (2.20 and 1.38) $z=2.5$. 

The \HIRAC\ colors are perturbed in all Voronoi bins simultaneously while maintaining a constant total flux. The \HIRAC\ color shift is calculated for each resolved element based on a normal distribution centered on the average of the four \HIRAC\ colors (where the IRAC flux is calculated by \autoref{eq:irac}) and a width of 1.0\,mag. The proposed perturbations are restricted to be physically plausible values by requiring the \HIRAC\ perturbations to fall within the range of the \HIRAC\ colors in the \citet{Bruzual03} stellar population synthesis model templates (constant SFH at the known redshift). The \HIRAC\ colors are perturbed for all but one Voronoi bin in the galaxy. If the sum of the perturbed IRAC fluxes is less than the original total IRAC flux, then the remaining Voronoi bin is attributed the leftover IRAC flux. Otherwise, if the sum of the perturbed IRAC fluxes add up to be greater than the original total IRAC flux, then all of the perturbations are repeated. The ``leftover'' bin often appears as an outlier in our results, but it does not alter the conclusions made here. An individual galaxy is perturbed $n$ times, where $n$ is the number of Voronoi bins in the galaxy, such that for each iteration a different Voronoi bin receives the leftover IRAC flux. In summary, for a given galaxy with $n$ Voronoi bins, the entire resolved distribution is perturbed $n$ times with every iteration having the same total flux. Finally, we use the perturbed IRAC fluxes to repeat the integrated and resolved SED modeling.

In \autoref{fig:perturb_compare}, the SED-derived properties are compared between the original results with constant \HIRAC\ colors and the perturbed solutions. Regardless of redshift, we find that a perturbed Voronoi bin (black points) on average (green circles) has the same properties as the case of constant \HIRAC\ colors (x-axis values), and by extension maintains its intrinsic color. \autoref{fig:perturb_ex} shows an example (GOODS-S\,36093, $z=2.62$) of how the resolved reddening and stellar population ages changes with each perturbation of the \HIRAC\ colors, further demonstrating that the colors do not vary significantly from bin-to-bin. While the Voronoi bin that received the leftover IRAC flux is occasionally seen as an outlier, the average \EBVs\ or stellar population age (values in the bottom right corner of each panel in \autoref{fig:perturb_ex}) remains consistent. In general, the average of the resolved distribution for all SED-derived properties is comparable with the average of the original distribution (light blue stars in \autoref{fig:perturb_compare}). Therefore, we conclude that we are not unduly constraining the resolved stellar ages or reddening by simplifying the IRAC flux distribution through our use of \autoref{eq:irac}.

\bsp
% MOVE LASTPAGE IF NECESSARY
%\label{lastpage}
\end{document}